\RequirePackage{fix-cm}
\documentclass[twocolumn]{svjour3}          %
\smartqed  %
\usepackage{graphicx}
\usepackage[hidelinks, breaklinks]{hyperref}

\journalname{}

\makeatletter
\def\makeheadbox{}
\makeatother

\usepackage{subcaption}

\usepackage{booktabs}  %
\usepackage{dcolumn}   %
\usepackage{multirow}  %
\usepackage{rotating}
\usepackage{xspace}
\usepackage{hyphenat}  %
\usepackage{xurl} %

\usepackage[table,dvipsnames]{xcolor}
\usepackage{enumerate}

\usepackage{array}    %

\usepackage{listings}
\definecolor{maroon}{cmyk}{0, 0.87, 0.68, 0.32}
\definecolor{halfgray}{gray}{0.55}
\definecolor{ipython_frame}{RGB}{207, 207, 207}
\definecolor{ipython_bg}{RGB}{247, 247, 247}
\definecolor{ipython_red}{RGB}{186, 33, 33}
\definecolor{ipython_green}{RGB}{0, 128, 0}
\definecolor{ipython_cyan}{RGB}{64, 128, 128}
\definecolor{ipython_purple}{RGB}{170, 34, 255}

\usepackage{algpseudocode}
\algrenewcomment[1]{\hfill// #1}%
\algnotext{EndFunction}
\algnotext{EndFor}
\algnotext{EndIf}

\usepackage{pifont}
\usepackage{tikz}
\usetikzlibrary{positioning, arrows.meta, backgrounds}
\usepackage{amsmath,amscd}

\usepackage{amssymb}

\usepackage[noeka]{mathrmletter}

\newif\ifextended
\newif\iflongbatching
\newif\ifsubmission
\newif\ifelementary

\ifx\buildextended\undefined
\else
    \extendedtrue
\fi

\ifx\noeditingmarks\undefined
   \newcommand{\pgwrapper}[3]{\begingroup \color{#1} #2: #3 \endgroup}
   \newcommand{\pgwrapperb}[1]{\textbf{#1}}
\else
   \newcommand{\pgwrapperb}[1]{}
   \newcommand{\pgwrapper}[3]{}
\fi

\newcommand{\sys}{Causal\-Mesh\xspace}

\newcommand{\globalmet}{globally available\xspace}
\newcommand{\Globalmet}{Globally Available\xspace}
\newcommand{\initial}{\noindent\textbf{Initial State:\xspace}}
\newcommand{\inductive}{\noindent\textbf{Inductive Step:\xspace}}

\newtheorem{fact}{Fact}

\def\compactify{\itemsep=0in \topsep=2pt \parsep=0.00in \partopsep=0pt
\leftmargin=2em}
\let\latexusecounter=\usecounter

  {\begin{list}{\labelitemi}{\itemsep3pt \topsep3pt \parsep0.00in
  \partopsep=3pt \leftmargin1.2em}}%
  {\end{list}}
\newenvironment{myitemize2}%
  {\begin{list}{\textbullet}{\itemsep1pt \topsep2pt \parsep0.00in
   \partopsep=1pt \leftmargin1.2em}}%
  {\end{list}}
  {\begin{list}{\labelitemi}{\itemsep2pt \topsep2pt \parsep0.00in
  \partopsep=0pt \leftmargin1.2em}}%
  {\end{list}}
  {\begin{list}{\threequartdash}{\itemsep3pt \topsep3pt \parsep0.00in
  \partopsep=3pt \leftmargin1.5em}}%
  {\end{list}}

\newenvironment{myenumerate2}
  {%
   \begin{enumerate}}
  {\end{enumerate}}

\newcommand{\topheading}[1]{\noindent\textbf{#1.}}
\newcommand{\heading}[1]{\vspace{1ex}\noindent\textbf{#1.}}
\newcommand{\weakheading}[1]{\vspace{1ex}\noindent\textit{#1:}}

\definecolor{codegreen}{HTML}{44732b}
\definecolor{codeblue}{HTML}{417ad2}
\definecolor{codered}{HTML}{911c0e}
\definecolor{codecomment}{HTML}{4e808c}

\lstdefinestyle{mystyle}{
    commentstyle=\color{codecomment},
    keywordstyle=\color{codegreen},
    stringstyle=\color{codered},
    numberstyle=\ttfamily\footnotesize\color{gray}\bfseries,
    basicstyle=\ttfamily\fontsize{7.3}{10}\selectfont\bfseries,
    breakatwhitespace=false,         
    breaklines=true,                 
    captionpos=b,                    
    keepspaces=true,                 
    numbers=left,                    
    numbersep=5pt,                  
    showspaces=false,                
    showstringspaces=false,
    showtabs=false,                  
    tabsize=2
}

\lstdefinelanguage{PythonPlus}[]{Python}{
  classoffset=1,
  morekeywords=[1]{,as,assert,nonlocal,with,yield,self,True,False,None,var} %
  morekeywords=[2]{,__init__,__add__,__mul__,__div__,__sub__,__call__,__getitem__,__setitem__,__eq__,__ne__,__nonzero__,__rmul__,__radd__,__repr__,__str__,__get__,__truediv__,__pow__,__name__,__future__,__all__,}, %
  morekeywords=[3]{,object,type,isinstance,copy,deepcopy,zip,enumerate,reversed,list,set,len,dict,tuple,range,xrange,append,execfile,real,imag,reduce,str,repr,}, %
  morekeywords=[4]{,Exception,NameError,IndexError,SyntaxError,TypeError,ValueError,OverflowError,ZeroDivisionError,}, %
  morekeywords=[5]{,ode,fsolve,sqrt,exp,sin,cos,arctan,arctan2,arccos,pi, array,norm,solve,dot,arange,isscalar,max,sum,flatten,shape,reshape,find,any,all,abs,plot,linspace,legend,quad,polyval,polyfit,hstack,concatenate,vstack,column_stack,empty,zeros,ones,rand,vander,grid,pcolor,eig,eigs,eigvals,svd,qr,tan,det,logspace,roll,min,mean,cumsum,cumprod,diff,vectorize,lstsq,cla,eye,xlabel,ylabel,squeeze,}, %
  keywordstyle=\color{codered},
  moredelim=[is][\color{codeblue}]{|}{|}
}
\lstdefinelanguage{DafnyLike}{
  morekeywords=[1]{predicate,var,requires,ensures,returns,true,false,and,or,if,then,else,match,case,return,while,break,assert,assume,invariant,forall,exists,ghost,reads,modifies,is,seq},
  keywordstyle=[1]\color{codegreen}\bfseries,
  morekeywords=[2]{ClientRead,integrate}, %
  keywordstyle=[2]\color{codeblue}\bfseries,
  morekeywords=[3]{Read,Read_Reply,State,Msg}, %
  keywordstyle=[3]\color{brown}\bfseries,
  sensitive=true,
  morecomment=[l]{\#},
  morestring=[b]",
}

\lstdefinestyle{dafnystyle}{
  language=DafnyLike,
  basicstyle=\ttfamily\fontsize{7.3}{10}\selectfont\bfseries,
  commentstyle=\color{codecomment},
  numberstyle=\ttfamily\footnotesize\color{gray}\bfseries,
  stringstyle=\color{codered},
  breaklines=true,
  breakatwhitespace=false,
  showstringspaces=false,
  showtabs=false,
  tabsize=2,
  numbers=left,
  numbersep=5pt,
  xleftmargin=15pt,
  keepspaces=true,
  captionpos=b
}

\usepackage{etoolbox}
\makeatletter
\patchcmd{\@maketitle}
  {\enskip{\boldmath$\cdot$}\enskip}
  {\,{\boldmath$\cdot$}\,}
  {}{}
\makeatother

\makeatletter
\AtBeginDocument{\@fleqnfalse}
\makeatother

\usepackage{fontawesome5}

\begin{document}

\title{CausalMesh: A Formally Verified Causally Consistent Distributed Cache with Support for Client Migration}
\author{Haoran~Zhang         \and
        Zihao~Zhang \and
        Shuai~Mu \and
        Sebastian~Angel \and
        Vincent~Liu 
}

\institute{Haoran Zhang \at
              University of Pennsylvania \\
              Lead author of the protocol and the implementation \\
              \email{haoranz@seas.upenn.edu}           %
           \and
           Zihao Zhang \at
              Stony Brook University \\
              Lead author of the formal verification effort \\
              \email{zihao.zhang@stonybrook.edu} 
           \and
           Shuai Mu \at
              Stony Brook University \\
              \email{shuai@cs.stonybrook.edu} 
           \and
           Sebastian Angel \at
              University of Pennsylvania \\
              \email{sebastian.angel@cis.upenn.edu} 
           \and
           Vincent Liu \at
              University of Pennsylvania \\
              \email{liuv@seas.upenn.edu} 
}

\date{} %

\maketitle

\begin{abstract}
Cloud applications often insert a caching lay\-er in front of a database in order to reduce I/O latency and improve throughput. 
One complication occurs when a client fetches some data from one cache node, then migrates to another (e.g., due to failures, load balancing, or client mobility), where it fetches the remaining data.
If the data in the cache nodes is inconsistent, the client could observe states that undermine the application's correctness.

One example of a situation where this is common is stateful serverless workflows, which consist of multiple serverless functions that access state in a remote database. 
In serverless, functions in the same workflow may be scheduled to different nodes with different caches, resulting in the migration pattern described above---the same client (the workflow) reads some data from one cache and other data from another.

To address this issue, this paper presents \sys{}, a novel approach to causally consistent distribu\-ted caching in environments where computations may migrate between machines.
\sys{} is the first cache system to support coordination-free, abort-free read/write operations and read transactions when clients migrate across multiple servers.
\sys{} also supports read-write transactional causal consistency in the presence of client migration, but at the cost of abort-freedom.

Our experimental evaluation shows that \sys{} has lower latency and higher throughput than existing proposals.
Finally, we have formally verified the correctness of \sys's protocol in Dafny.

\end{abstract}

\section{Introduction}\label{s:intro}

Large-scale applications often rely on remote databases
to manage state.
Given that accessing remote databases is expensive (e.g.,
10--20\,ms to read or write to DynamoDB), developers frequently
add a cache layer between their applications and the database to
improve I/O latency~\cite{hellerstein19serverless,pu2019shuffling}.
As systems scale, these caches are often distributed and
partitioned across many nodes, bringing the data closer to the
computations but raising the challenge of maintaining consistency.
Proposals here generally include having (i) a large cache or
multiple caches with a cache coherence protocol, which provides
strong consistency but does not scale, or (ii) a cluster of
distributed caches, such as Amazon DynamoDB Accelerator (DAX)~\cite{daxconsistency},
that scales well but provides only weak (eventual) consistency.

The issue is that weak consistency is problematic when a client makes a
sequence of requests as part of a single logical session to
different cache servers (i.e., the client migrates from one
cache server to another).
This migration can happen because the client explicitly consults a different cache server, because of natural handover in a mobile wireless network, because the original cache fails and the new cache is acting as a backup, or because a cloud provider chooses to schedule the client on a different machine.

To see the impact of weak consistency, imagine a social media site where a client session executes two consecutive operations that access overlapping state: the first operation blocks a creator or flags content as inappropriate, and the second operation refreshes the user's feed~\cite{cooper2008pnuts,lloyd2011cops}. If these two operations are served by different cache servers, they may observe different states. In such a case, the client may fail to read its own writes (i.e., the second operation does not observe the effect of the first), and the user may continue to see unsafe or blocked content. This violates basic session consistency and illustrates why weakly consistent caches can make building correct distributed applications challenging.

One place where this type of client migration is not just possible, but actually extremely common is in \emph{serverless computing}.
Serverless functions allow clients to run their applications on
cloud providers without needing to manage or operate servers,
load-balance requests across VMs/containers, scale resources up
or down based on load, or handle failures.
In serverless architectures, developers implement their logic
using \emph{workflows}, which are directed graphs of stateless
functions.
Because these functions are stateless, the serverless runtime
can schedule them on arbitrary machines to optimize resource
utilization.
When developers want these functions to maintain state across
invocations, serverless functions connect to a remote database,
such as DynamoDB, often with an additional distributed cache
layer, such as DAX.
Consequently, different functions within the same workflow
execution may run on different machines and connect to different
cache nodes, leading to anomalies inherent to eventually
consistent caches.

\begin{figure}[t]
  \centering
  \begin{subfigure}[b]{0.25\columnwidth}
    \centering
    \includegraphics[width=\textwidth]{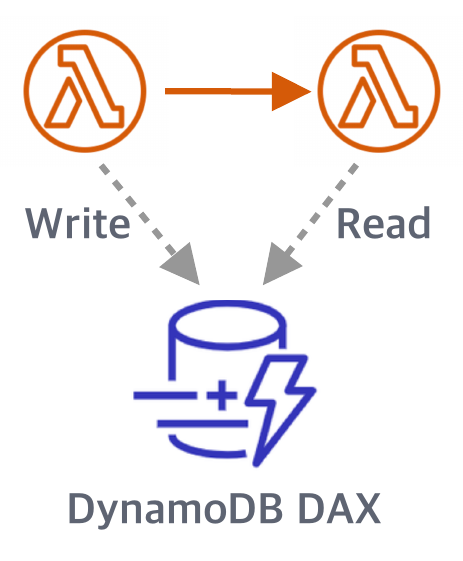}
  \end{subfigure}
  \begin{subfigure}[b]{0.5\columnwidth}
    \centering
    \includegraphics[width=\columnwidth]{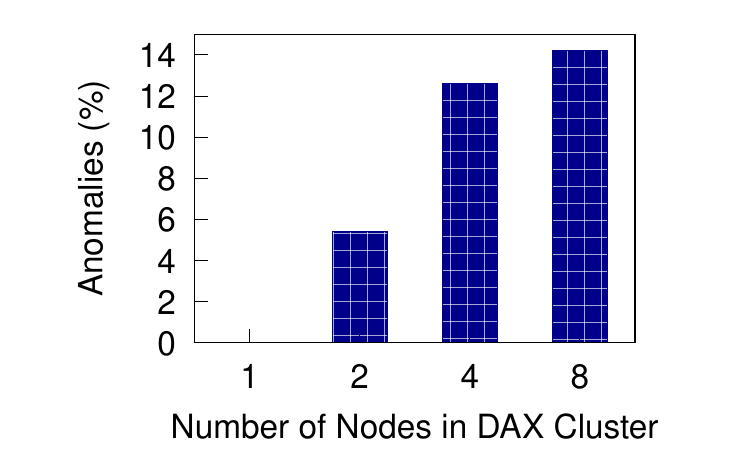}
  \end{subfigure}
  \caption{Anomalies rate of a two-function workflow,
  where the second function reads the data written by the first. The workflow runs on AWS Lambda
  and using DynamoDB Accelerator (DAX) as the cache.
  There are no anomalies when utilizing a single cache node, but it lacks scalability. }%
  \label{fig:dax}
\end{figure}

To better characterize this issue, we implement a minimal serverless workflow on AWS Lambda and DynamoDB DAX.
The workflow consists of two serverless functions that access the same state: the first writes to the state, and the second reads from it.
As shown in Figure~\ref{fig:dax}, we observe that in this simple example, the anomaly probability can be as high as 14.2\% when there are 8 cache nodes in DAX.

Recent works, in particular HydroCache~\cite{wu2020hydrocache} and FaaSTCC~\cite{lykhenko2021faastcc}, aim to address this issue by introducing a \emph{causal cache}: a set of caches that collectively guarantee causal consistency.
Both HydroCache and FaaSTCC provide transactional causal consistency~\cite{akkoorath2016cure,lloyd2011cops}, which they adopt from traditional causal databases.
The main technical challenge addressed by them and other prior works~\cite{zawirski2015writefast,mehdi2017occult} is dealing with \emph{client migration}: a client (or a serverless workflow) can access one cache during one operation, and then a completely different cache in another operation within the same session.
 
All prior works handle client migration by introducing expensive coordination and aborts that significantly reduce the benefits of introducing a cache in the first place.
In particular, HydroCache requires cache servers to coordinate before execution to fetch the necessary versions of data items. 
It also requires aborting and retrying the entire session/workflow when transactions fail to commit.
Both can introduce significant overhead to applications.
   
To improve the performance of distributed caching and ensure that applications under client migration work as intended, we present \emph{\sys}, a novel distributed cache that supports client migration securely and efficiently.
\sys{} has several features:

\begin{myenumerate2}
  \item \emph{Per-session causal consistency.} \sys{} ensures that any data accessed by a client session (e.g., a serverless workflow) observes the effects of prior operations in the same session, even if they run on different servers and access different caches.
  \item \emph{Coordination-free reads and writes.} In \sys{}, a cache never has to synchronize with peers to process an operation.
  \item \emph{No aborts.} All operations in a session that use \sys{} always read from a causally consistent snapshot, so they never need to abort due to inconsistencies.
  \item \emph{High throughput and low latency.} \sys{} achieves high throughput and its latency is low and stays nearly constant as we vary the number of caches.
\end{myenumerate2}

Achieving coordination-free caching with causal consistency under client migration is challenging: when a client writes to one server and then migrates to another, the new server may not yet have the write, and serving a read without coordination risks violating causal consistency.
\sys{}'s core insight is to decouple reads and writes using a \emph{dual cache}~(\S\ref{s:dualcache}): each server maintains an \emph{Inconsistent cache} (I-cache) for accepting writes, and a \emph{Consistent cache} (C-cache) for serving reads.
A write becomes visible only after it is \emph{globally causally consistent}.
That is, the write and all of its transitive dependencies have been observed by every server.

Following the lifecycle of a write illustrates how the mechanisms fit together.
When a client writes to a cache server, the write is stored in the server's I-cache and acknowledged immediately, without coordination (\S\ref{s:writepath}).
The write is then asynchronously \emph{propagated} along a chain that connects all servers~(\S\ref{s:propagation}).
Once propagation completes, the write is safe to be \emph{integrated}: moved from I-cache to C-cache, along with all of its transitive dependencies~(\S\ref{s:pull}).
Reads are served entirely from the C-cache, which is always a causally consistent snapshot, so they require no coordination with other servers~(\S\ref{s:readpath}).
  
Ensuring the correctness of the above protocol is non-trivial.
Our formal verification effort in Dafny uncovered a subtle but critical causal consistency violation in the original design~\cite{CausalMesh}, where we had used a single round of chain propagation.
The bug arises when a write is integrated into the C-cache before all servers have observed its transitive dependencies, allowing migrating clients to read causally inconsistent state.
This violation was not detected in our original work, despite the fact that we had performed bounded model checking with TLA+ and TLC.
In particular, we missed a case that only shows up once you have a sufficient number of clients and cache servers. 
We redesigned the propagation protocol to use two rounds~(\S\ref{s:propagation}), and our Dafny proof establishes correctness for arbitrary numbers of clients and cache servers~(\S\ref{sec:verification}).

To make the benefits of \sys{} broadly applicable, we build a library that exposes an intuitive interface similar to that of a traditional key-value store~(\S\ref{s:clientlib}).
Developers can use this library to build their applications.
We also describe a variant of \sys, \sys{}-TCC~(\S\ref{s:tcc}), that provides support for arbitrary read/write transactions across multiple operations within a session (e.g., across multiple serverless functions), although this comes at the cost of losing the abort-free property.

We implement \sys{} and \sys{}-TCC on top of Nightcore~\cite{jia2021nightcore}, a serverless runtime platform.
We then use the key-value store interface provided by \sys's client library to write several microbenchmarks~(\S\ref{s:eval:micro}) and real-world applications consisting of workflows with 13 serverless functions~(\S\ref{s:eval:real}) to evaluate the performance.

To put our results in context, we compare \sys{} and \sys{}-TCC with HydroCache~\cite{wu2020hydrocache} and FaaSTCC~\cite{lykhenko2021faastcc}, recent caches for serverless workflows that aim to play a similar role. 
In a nutshell, \sys{} is significantly faster: we observe an up to 59\% reduction in median latency, up to a 97\% reduction in tail latency and 1.3--2$\times$ higher throughput.
Furthermore, caches in \sys{} do not need to coordinate with other caches or abort (whereas HydroCache and FaaSTCC must do one of the two). 
When we extend the comparison to the transactional variant of \sys, \sys{}-TCC, we observe that \sys{}-TCC still achieves 1.3--1.8$\times$ higher throughput than, and comparable latency to, HydroCache and FaaSTCC.

In summary, the contributions of this work are:
\begin{myenumerate2}

\item \sys, a cache system that provides causal+ consistency. 
To our knowledge, \sys{} is the first general causal cache system that supports coor\-dination-free reads and writes in the presence of client migration.
\item A lock- and coordination-free read transaction protocol 
  that allows developers to get a causally consistent view across multiple 
  keys within a single operation (e.g., a serverless function).
\item \sys{}-TCC, an extension of \sys{} that supports transactional causal consistency across multiple operations in a session.
\item The implementation and experimental evaluation of \sys{} that demonstrates its low latency and high throughput.
\item A rigorous machine-checked proof of correctness in Dafny~\cite{leino2010dafny} that establishes that \sys{} preserves causal consistency in arbitrary system configurations.

\end{myenumerate2}

\section{Background and Goals}\label{s:bg}

We begin by providing context on serverless execution models as well as our 
  target consistency levels.
\subsection{Serverless Architecture}

When deploying a traditional, serverful application to the cloud, users allocate 
  VMs and deploy their software to the resulting instances.
While the cloud handles the management of the physical infrastructure,
  users remain responsible for many tasks before their applications can execute, 
  e.g., requesting a batch of VMs from the cloud provider, 
  specifying their resource profiles, choosing their base VM images, 
  setting permissions/firewall rules, deploying dependencies, 
  and monitoring the application as it runs, among others.

Serverless computing promises to free users from all of the above concerns.
Instead, users supply the cloud provider with a function that executes their 
  application logic, and the provider handles all provisioning, scaling, 
  load balancing, and management of the execution instances.
The functions can even be composed into \emph{workflows}, which are graphs 
    of serverless functions that collectively perform the logic of an application.
Two aspects of this architecture are particularly salient to the design and 
  necessity of \sys:

\heading{(1) Provisioning and scheduling}
Unlike in traditional execution environments, one of the core responsibilities
  of cloud providers in serverless is managing function workers and 
  assigning requests to those workers, all of which are done out of the view of users.

At a high level, the typical strategy operates as follows.
When a request for a function arrives and finds that all existing instances of that function
  are busy, the provider will deploy a new instance of the function to handle the
  request, i.e., a cold start.
After handling the request, the instance will be kept warm (provisioned) for 
  some time before being reclaimed---up to 1\,hr in the case of AWS Lambda~\cite{lambda-warm-start}.
Requests are generally handled in FIFO order and routed to random instances
  among the set of unsaturated, pre-warmed instances when possible.

For a workflow that has a few functions, each function can be assigned to a different worker.
We say a workflow \emph{migrates} to a new worker when a function in the workflow is allocated to a different worker than its predecessor. 
Note that a workflow can migrate to multiple workers concurrently 
  if it has a fan-out structure.

In reality, the workers that execute a workflow are typically located close to each other, e.g., in the same data
  center or availability zone, because a cluster often defines the management boundary for workloads.
  Once a workload is deployed to a cluster, it is typically not moved to another cluster because each
  cluster usually has its own isolated control plane~\cite{tang2020twine}. In AWS Lambda, to improve cache locality,
  enable connection re-use, and amortize the costs of moving and loading customer code,
  events for a single function are sticky-routed to as few workers as possible~\cite{agache2020firecracker}.

\heading{(2) State management}
A side effect of the above approach is that users must carefully manage any 
  state that should persist across function executions, as the number of 
  underlying instances and the routing of requests to instances is opaque to users.
There is no guarantee or method to enforce that two requests will be executed in 
  the same instance, whether the requests are for the same function or
  different functions in the same workflow.
For these so-called Stateful Serverless Functions (SSFs), external storage 
  services, e.g., relational databases or key-value stores, are standard 
  solutions for persisting application state.
Of course, access to these remote services can incur high latency and block 
  critical path execution.

\heading{(3) Caching}
To reduce the latency of accessing remote storage services, a cluster of cache nodes is deployed between the application and the remote storage. 
Taking Amazon's DynamoDB Accelerator (DAX) as an example, 
using the write-through mode, a write
  request is first directed to the primary cache and then replicated to other cache nodes.
This replication is eventually consistent and can take up to 1 second to complete.
Consequently, two clients may obtain different values when accessing
  the same key from the same DAX cluster, depending on the node that each client accesses.

\subsection{Consistency Goals}\label{s:bg:goals}
One potential solution to the high overhead (particularly high latency) of 
  state management is to cache the remote state at each provisioned instance, 
  allowing functions to access the state immediately if the data is in the cache.
Unfortunately, to maintain consistency across an entire workflow, traditional caches 
  generally either need to block and confirm that they have the latest state by 
  synchronizing with other caches, or they must proceed speculatively but then abort 
  if an inconsistency is ever detected (as is the case in 
  systems like HydroCache~\cite{wu2020hydrocache} and classic cache coherency protocols).
This results in higher latency, particularly at the tail.
Another approach altogether is to ignore strong consistency in favor of weaker 
  guarantees (as is the case in AWS's DAX service~\cite{daxconsistency}), but 
  as we alluded to in the introduction, writing serverless workflows with 
  weak consistency is very challenging.
To strike a balance between excellent performance and meaningful consistency 
  semantics, we settle on \emph{causal+ consistency} (CC+).
Recent work~\cite{mahajan2011consistency} has shown that no model stronger than
  causal consistency is achievable with high availability, making it ideal
  for our coordination-free goal. CC+ can be summarized as:

\begin{myenumerate2}
\item\textbf{Client-side dependency.} If an operation is issued after another
  operation is completed by the same client, the latter operation must observe the
  effects of the former. In the context of serverless, a client is a
  workflow.

\item\textbf{Read-write dependency.} If a read operation reads the effect of a
  write operation, then we say there is a read-write dependency between the two
  operations. This clause itself is not a guarantee, but together with the
  following rule, it restricts the system's behaviors. 

\item\textbf{Transitive dependency}. The above two dependencies are transitive.
  Transitive dependencies must be respected by the execution. For example, if a
  read operation transitively depends on a write operation, the write must be
  reflected in the read results.  

\item\textbf{State convergence}. Different replicas of the same data will
  eventually converge to the same state. This is also known as causal+
  in the context of causal consistency. 
\end{myenumerate2}

Providing causal consistency in the cache can greatly simplify programming in 
  serverless workflows and make them less error-prone. 
A simple example is that it can avoid the anomaly discussed in \S\ref{s:intro}.  
In a more complex example, consider a serverless workflow that implements a Twitter-like 
  social media service.
This example was previously implemented in serverless by Beldi~\cite{Beldi} 
  and was ported from the microservice library DeathStarBench~\cite{deathstarbench}. 
When Alice replies to Bob's post, a serverless function will store the reply in the database's reply table and notification table; it also stores the id of the reply in the database's post table as foreign keys. When Bob receives the notification and interacts with this serverless application, a serverless function will fetch the post content and all its replies to render the page. 
There is a dependency between the notification and the post's replies, and without 
  causal consistency, when the serverless function returns the page to 
  Bob with the rendered post, it might not contain the reply that triggers the notification.
Another common example includes applications whereby a user sets permission (e.g.,
  removes a user from an access control list), and then posts a sensitive file. 
Without causal consistency, the removed user may see the sensitive file~\cite{cooper2008pnuts,lloyd2011cops,pang2019zanzibar}.

\subsection{Challenges}
Serverless environments present a unique challenge to maintaining causal 
  consistency.
One key reason is the extremely fine-grained resource provisioning and
  autoscaling that is central to the serverless approach---low latency is 
  achieved by routing functions preferentially to instances that are provisioned 
  and available, even if they access a cold cache.
Spinning up a new instance on the local machine or allowing longer queues for local execution might improve cache hit rates (e.g., using a technique like~\cite{abdi2023palette}), but without extremely restrictive scheduling policies, there is no guarantee of correctness.
It also potentially comes at the cost of the provider 
  resources that are not included in current pricing models.

Traditional causal cache systems like Bolt-on~\cite{bailis2013bolt} cannot be directly applied in this scenario.
Bolt-on uses a background thread that subscribes to the database and periodically
  merges new updates into the cache. 
In a \textit{sticky} setup like Bolt-on, where the client always communicates 
  with the same cache server, this approach is efficient and correct because 
  the data in the cache are monotonic and never roll back, ensuring clients never read an older 
  version than the one they previously read. 
However, in a serverless environment, different functions within a workflow could be scheduled to different instances and interact with different cache servers.
  Even though each cache is monotonic, a cache server might provide a version older than one accessed from another cache server.
This means that clients may fail to find a version they previously read if they
  migrate to another server.

\section{\sys Overview}\label{s:overview}\label{s:design}
The goal of \sys{} is to provide a high-performance, resilient, and causally consistent distributed cache service that supports client migration.
While \sys's design is general, we will describe it in the context of a serverless platform to make some of the more nuanced concepts very concrete.

\subsection{Architecture}
\begin{figure}
  \centering
  \includegraphics[trim={4cm 1cm 4cm 2cm}, clip, width=\columnwidth]{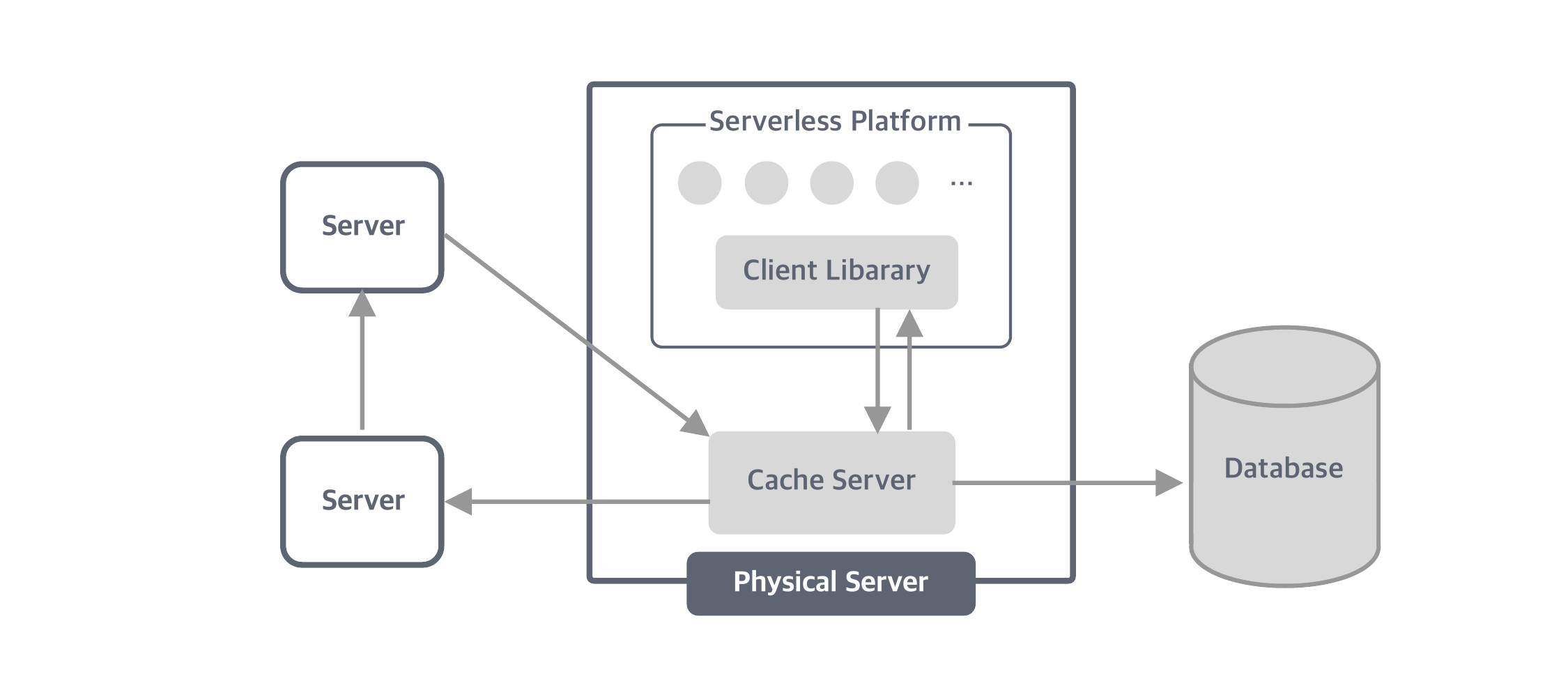}
  \caption{Architecture of \sys{} in a three-server setup.}%
  \label{fig:arch}
\end{figure}

Figure~\ref{fig:arch} illustrates the architecture of \sys{}. 
The architecture consists of four components:

\heading{Serverless Platform} The serverless platform acts as the runtime 
  environment for user applications. It orchestrates the workflow and dispatches functions to available workers.

\heading{Databases} The database stores user data.
\sys{} supports any database, 
  as long as the database allows custom conflict resolution policies to resolve concurrent updates (e.g., Azure CosmosDB~\cite{cosmosdb}, Couchbase~\cite{couchbase}, and MongoDB~\cite{mongodb}).

\heading{\sys{}} \sys{} is a middleware that sits between the serverless platform 
  and the backend databases. It contains two components, cache servers and a client library. The user functions interact with the cache servers using the client library. Cache servers communicate 
  with each other via remote procedure calls (RPCs). The messages between cache servers follow a FIFO order but can experience arbitrary delays.
\sys{} plays a similar role to DynamoDB DAX or HydroCache.
In our setup, \sys{}'s cache server runs in a 1:1 correspondence with 
  physical machines, each physical machine runs a serverless worker and a cache server.
All requests are routed to the cache server on the same machine. This setup provides the best locality.
However, other configurations are also possible. For example, machines in the same rack may be assigned to the same cache server. Cache servers are managed and configured by a fault-tolerant coordinator (e.g. Zookeeper~\cite{hunt2010zookeeper}).

\medskip

Using the above components, the journey of a stateful serverless workflow proceeds as follows:

\begin{myenumerate2}
\item The workflow is triggered by an event, e.g., a request from a browser or some service arriving at a gateway.
\item The scheduler in the serverless platform dispatches the first function in the workflow to a worker 
  machine based on resource usage, hardware requirements, and other factors.
\item The function accesses state by using \sys's client library to communicate with the cache  
  on the same machine. 
\item After the function finishes, the scheduler gathers the result before dispatching the subsequent function or functions (in the context of a fan-out workflow) to potentially different worker machine(s) than the previous one.
\item Repeat until the workflow is complete.
\end{myenumerate2}

\subsection{CC+ in \sys{}}

\sys{} uses vector clocks and dependencies to enforce CC+. 

\heading{Vector Clocks (VC)} Vector clocks~\cite{singhal1992efficient} are 
  used to identify different versions of an object and capture the 
  happens-before relation~\cite{lamport2019time} between them. 
We use version and vector clock interchangeably in the paper. 
A vector clock $VC$ is a set of $\langle$server id, timestamp$\rangle$ pairs;
  each server maintains its corresponding timestamp and increments it as needed.
  For simplicity, we assume that given $N$ servers, each server is assigned an id
  from $0$ to $N-1$ so that a VC can be represented using a list of timestamps
  where $VC[i]$ is the timestamp of server $i$.

We define the union of two VCs, $VC_1 \cup VC_2$, as their element-wise 
  maximum, e.g., $[1, 0] \cup [0, 1] = [1, 1]$. 
By using vector clocks, we can implement a custom conflict resolution policy 
  to ensure state convergence in CC+. 
Informally, new versions overwrite old versions; if two versions are 
  concurrent, we merge the vector clocks and pick one of the values as 
  the new value in a deterministic way.
In the implementation, we break ties by picking the value of the larger version by lexicographical ordering.

\noindent

\heading{Dependencies (deps)} 
Dependencies are used to track causal relationships across different keys.
They are stored as a map from a key to the vector clocks of the writes that it depends on.
\setlength{\belowdisplayskip}{3pt} \setlength{\belowdisplayshortskip}{3pt}
\setlength{\abovedisplayskip}{3pt} \setlength{\abovedisplayshortskip}{3pt}
\[
    \begin{array}{rcl}
        deps &:=& \{Key \mapsto VC \} \\
    \end{array}
\]
To reduce the size of metadata, it only contains the nearest dependencies, meaning that if $x \to y$ ($x$ happens 
  before $y$, $y$ depends on $x$) and $y \to z$, $z$'s dependency will contain $y$ but not $x$.  Dependencies can be merged using the same mechanism as vector clocks.

\section{\sys{} Protocol}\label{s:protocol}

\newcommand{\black}{I-cache\xspace}
\newcommand{\white}{C-cache\xspace}

This section describes the internal mechanisms of \sys{}.
We first introduce the dual cache data structure that enables
  coordination-free operation~(\S\ref{s:dualcache}), followed by
  the client library and APIs~(\S\ref{s:clientlib}).
We then describe the write path~(\S\ref{s:writepath}),
  propagation and integration~(\S\ref{s:propagation}), the read
  path~(\S\ref{s:readpath}), read
  transactions~(\S\ref{s:readtxn}), and conclude with an intuitive
  argument for correctness~(\S\ref{s:cc}).

\subsection{The Dual Cache}\label{s:dualcache}

\begin{figure*}[tb]
{
\resizebox{\textwidth}{!}
{\begin{tabular}{l|l}
\toprule
\textbf{\sys{} Server API} & \textbf{Description} \\
\midrule
\texttt{ClientRead(key, deps)} $\to$ \texttt{value, vc} & client's read request with a key and its dependencies, return the value and version. \\
\texttt{ClientWrite(key, value, deps, local)} $\to$ \texttt{vc} & client's write request with the key, value, dependencies and the client's own writes, return the version. \\
\texttt{ClientReadTxn(keys, deps)} $\to$ \texttt{values, vcs} 
 & client's read transaction request with keys and their dependencies, return values and their versions. \\
\texttt{ServerWrite(key, value, vc, deps)} & write request from another server with the key, value, version and dependencies. \\
\bottomrule
\end{tabular}
}
}
\vspace{1ex}
\caption{\sys{}'s Server APIs.
The first three APIs (\texttt{Client*}) are used in \sys{}'s client library.
\texttt{ServerWrite} is called by other \sys servers via RPC to propagate writes.
Note that users do not interact directly with these functions.
}
\label{fig:table:api}
\end{figure*}

\begin{figure}
\hrule
\begin{lstlisting}[language=PythonPlus]
# self is a client (workflow)
def |read|(self, key):
    # Send read request with deps to server
    value, vc = |ClientRead|(key, self.deps)
    # Track this read as a dependency
    self.deps.|merge|(key, vc)
    # Merge with local write if exists (RYW)
    if key in self.local:
        return |merge|(self.local[key],
               {value, vc}).value
    return value

def |write|(self, key, value):
    # Send write with deps and local to server
    vc = |ClientWrite|(key, value,
                     self.deps, self.local)
    # Record in local for RYW on migration
    self.local[key].|merge|(value, vc)

\end{lstlisting}
\hrule\vspace{2ex}

\caption{Pseudo-code for \sys{}'s Client Library.}%
\label{fig:alg-client}
\vspace{-8pt}
\end{figure}

A core component of \sys{} is its \textit{dual cache},
which enables coordination-free reads and writes.
Each cache server maintains an instance of this dual cache, which
  is essentially two subcaches, a \textit{Consistent cache}
  (\white{}), and an \textit{Inconsistent cache} (\black{}).
\[
    \begin{array}{rcl}
        \textit{\white{}} &:=& \{~Key \mapsto (~Value, VC~)~\} \\
        \textit{\black{}} &:=& \{~Key  \mapsto
          [~(~Value, VC, deps~)~]~\} \\
    \end{array}
\]
\white is a hash map from keys to values and their corresponding
  versions; it acts like a single-version key-value store.
All versions in \white{} are guaranteed to be synchronized on all
  cache servers and, therefore, visible to clients.
Because \white{} is always a causally consistent snapshot,
  read operations can be served from \white{} without consulting
  other servers.

\black, on the other hand, is a hash map from keys to a tuple
  (Value, VC, Deps).
It acts like a multi-version key-value store and stores versions
  that the cache server is unsure whether they have been
  synchronized to all servers.
As a result, \black is unsafe to reveal to clients, but
accepting writes into \black{} requires no coordination
  with other servers, so write operations return immediately.

The bridge between the two is \textit{integration}
  (\S\ref{s:pull}), a purely local procedure that moves versions
  from \black{} to \white{} once they have been propagated to all
  servers.
This separation is what makes \sys{} coordination-free: writes
  go to \black{} (fast, no coordination), reads come from
  \white{} (consistent, no coordination), and integration bridges
  the two once writes have been globally propagated.

The invariant maintained by the dual cache is that \white{} is
  always a \textit{strict causal cut}, or simply a cut.
Informally, this means that the dependencies for each write in the
  cut should either be in the cut or should happen before a write
  to the same key that is already in the cut.
The formal definition is as follows.

\begin{definition}[Strict Causal Cut]
A set of writes $S$ is a strict causal cut
  $\iff \forall x \in S, \forall y \in x.deps, \exists y' \in
  S~|~y.key = y'.key \land (y' = y \lor y \to y')$
\end{definition}

When evicting a key $k$, all keys that depend on it are also
  evicted so that C-cache remains a cut.

\subsection{Client Library and APIs}\label{s:clientlib}\label{s:api}

\sys{}'s client library offers an intuitive interface for
  developers, similar to a traditional key-value store:

{
\vspace{3pt}
\small
\begin{myenumerate2}
\item \texttt{Read(key)} $\to$ \texttt{value}
\item \texttt{Write(key, value)}
\item \texttt{ReadTxn(keys)} $\to$ \texttt{values}
\end{myenumerate2}
}

\noindent
The \texttt{ReadTxn} operation returns a causally consistent view
  of multiple keys within a single serverless function.
Developers use these three operations to build their applications;
  the consistency guarantees are handled transparently by the
  client library and the cache servers.

\begin{figure*}[t]
  \centering
  \includegraphics[width=.8\textwidth]{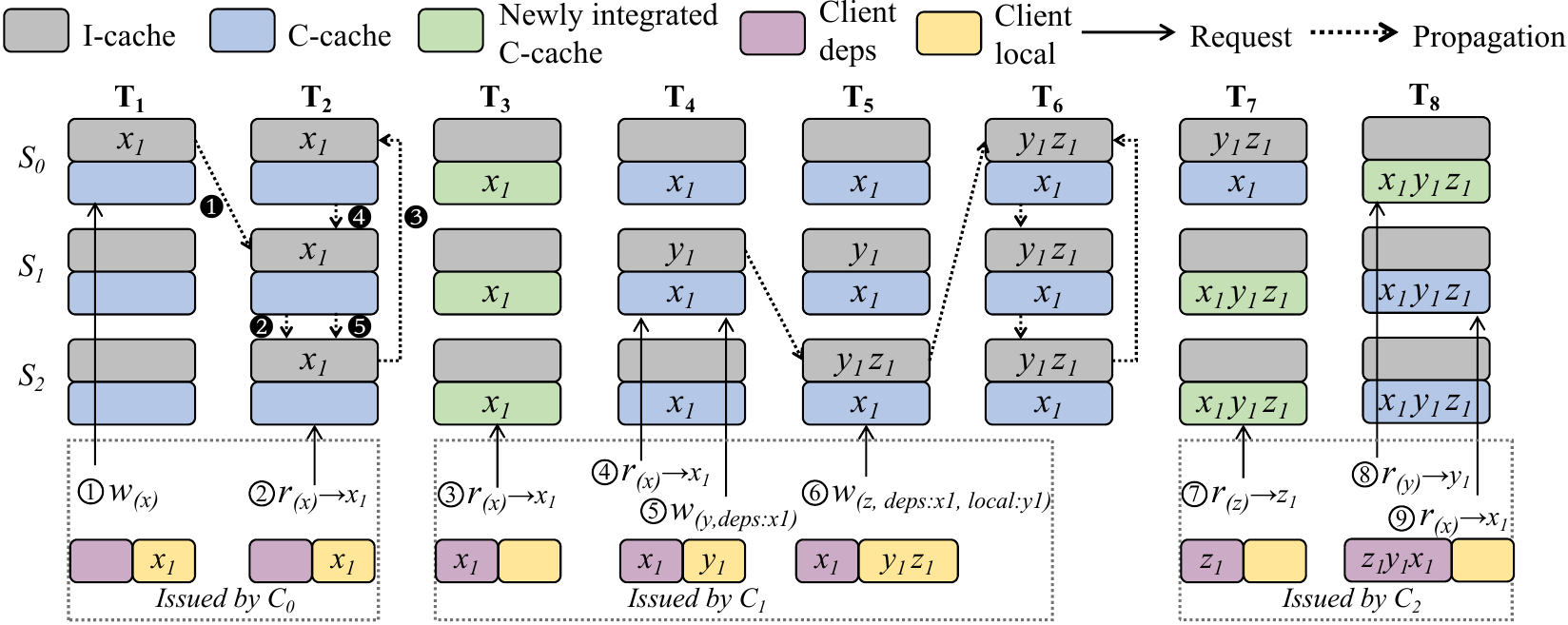}
  \caption{Running example illustrating \sys{}'s end-to-end
    operation with three clients $C_0$, $C_1$, $C_2$ and three
    servers $S_0$, $S_1$, $S_2$.
  Each column shows the I-cache (gray) and C-cache (blue) state of
    every server at that time step;
    green cells indicate versions newly integrated into C-cache.
  Below each client group, pink cells show the
    client's \textit{deps} and
    yellow cells show the client's
    \textit{local} at each stage.
  White-circled numbers (\ding{192}--\ding{200}) denote client
    operations; black-circled numbers (\ding{202}--\ding{206})
    denote propagation steps between servers.
  $C_0$ writes $x$ and reads it back via \textit{local}
    before propagation completes (Read Your Writes,
    steps~\ding{192}\ding{193}).
  $C_1$ reads $x$ on two different servers and observes the same
    version (Monotonic Reads,
    steps~\ding{194}\ding{195}), then writes $y$ with $x_1$ in
    \textit{deps} (Writes Follow Reads,
    step~\ding{196}), and migrates to $S_2$ to write $z$ while
    carrying $y_1$ in \textit{local}
    (Monotonic Writes, step~\ding{197}).
  $C_2$ reads $z$, then $y$, then $x$: the causal chain
    $x_1 \to y_1 \to z_1$ ensures that observing $z_1$ implies
    observing both $y_1$ and $x_1$
    (steps~\ding{198}--\ding{200}).
    }%
  \label{fig:running}
\end{figure*}

The client library translates each client operation into a
  corresponding server API call (Figure~\ref{fig:table:api}), which
  carries additional metadata.
To understand this metadata, we first describe the two pieces
  of state that the client library maintains throughout a
  workflow.

\heading{Client state: \textit{local} and \textit{deps}}
In serverless computing, a client refers to a workflow made up of
  multiple functions.
When a workflow starts, the client library creates two maps,
  \textit{local} and \textit{deps}.
These two maps track the client's own writes and read dependencies,
  respectively.
When the scheduler dispatches a subsequent function to a
  (potentially different) server, it passes \textit{local} and
  \textit{deps} along with the workflow context, so the new
  function's client library inherits the causal state of the
  previous one.
It is important to note that the \textit{deps} map does not contain
  the client's own writes, even though they are semantically part
  of the dependencies.
The \textit{local} map is what enables clients to read their own writes 
  even when they have not yet been
  propagated~(\S\ref{s:writepath}): the client library merges the
  server's response with its own write from \textit{local},
  guaranteeing that a client always observes its own effects.

Figure~\ref{fig:alg-client} shows the pseudo-code of the
  client library.
The detailed read and write protocols, which describe how the
  client library interacts with the cache server, are presented in
  the next three subsections.

\subsection{Write Path}\label{s:writepath}

\heading{Client side: issuing the write}
Each write in \sys{} (CC+) operates on a single key independently.
When a client invokes \texttt{Write(key, value)}, the client
  library wraps it into a \texttt{ClientWrite} request that
  attaches both the \textit{deps} and \textit{local} maps as
  metadata, and sends it to the cache server the client is
  connected to (typically co-located in a serverless environment).
For example, in Figure~\ref{fig:running}, step~\ding{196} shows
  $C_1$ writing $y$ to $S_1$ with $deps{:}x_1$, and
  step~\ding{197} shows $C_1$ writing $z$ to $S_2$ with
  $local{:}y_1$.

\heading{Server side: storing the write}
The server-side write processing is shown in the
  \texttt{ClientWrite} function of Figure~\ref{fig:alg-server}.
Each write is initially saved in the receiving server's \black{},
  as it only exists on that single server and is not yet ready to
  be made visible.
Upon saving the write to \black{}, \sys{} assigns it a version
  based on the server's global vector clock (\textit{GVC}).

Each cache server $S_i$ maintains its own \textit{GVC}, a vector
  clock where \textit{GVC}[$i$] counts the writes issued at $S_i$
  (used as a unique identifier for new writes) and
  \textit{GVC}[$j$] ($j \ne i$) tracks the largest version of
  server $S_j$ that $S_i$ has integrated into \white{}.
When receiving a write, $S_i$ increments \textit{GVC}[$i$] and
  assigns the resulting vector clock to the write.
For example, if $S_0$'s \textit{GVC} is $[7, 5, 2]$, a new write
  is assigned $[8, 5, 2]$.

After assigning a new version to a write, the cache server adds
  the write to \black{}, flushes it to the database, and returns
  the assigned vector clock to the client without waiting for any
  other server.
At this point, the new value is not yet visible to other clients.
To make the value visible, the server asynchronously sends the
  write to its successor in the propagation
  chain~(\S\ref{s:propagation}).

To reduce write latency, the cache server can optionally flush
  writes to the database asynchronously via a local persistent
  log.
In this mode, only the server holding the log has the full write
  history.
If that server crashes and cannot be recovered, the unflushed
  writes are lost.
This weakens durability but does not violate causal consistency.
A lost write was never propagated or flushed, so no other client
  has observed it.

\heading{Client side: recording the write}
Upon receiving the vector clock from the server, the client
  library records the write in its \textit{local} map.
Subsequent operations within the same workflow can use
  \textit{local} to ensure Read Your Writes
  (see~\S\ref{s:readpath}), and the next function in the workflow
  inherits the updated \textit{local} after migration.
From the client's perspective, the write is now complete.

\begin{figure}
\hrule
\begin{lstlisting}[language=PythonPlus]
# self is a cache server
def |ClientRead|(self, key, deps):
    self.|integrate|(deps)  # integrate deps
    return self.Consistent[key]


def |ClientWrite|(self, key, value, deps, local):
    # Merge local into deps
    for k, (v, vc, k_deps) in local.items():
        deps.|add|(k, vc)
    # Update server vc
    for k, vc in deps.items():
        self.vc.|merge|(vc)
    self.vc[self.id] += 1  # new version
    # Store in I-cache
    self.Inconsistent[key].|add|((
        self.vc, value, deps
    ))
    # Async propagation
    self.successor.|ServerWrite|(
        key, self.vc, value, deps
    )
    return self.vc  # ack

def |ServerWrite|(self, key, vc, value, deps):
    if self is tail:  # 2nd round done
        self.vc.|merge|(vc)
        self.|integrate|(deps)
        self.Consistent[key].|merge|(
          (vc, value, deps)
        )
    else:
        if not self.has_seen(key, vc):
            # Round 1: store in I-cache
            self.Inconsistent.|add|(
                (vc, value, deps)
            )
        # Round 2: skip store, just forward
        self.successor.|ServerWrite|(
            key, vc, value, deps
        )

\end{lstlisting}
\hrule\vspace{3ex}
\caption{Pseudo-code for \sys{}'s Server.}
\label{fig:alg-server}
\end{figure}

\subsection{Propagation and Integration}\label{s:propagation}\label{s:pull}

\begin{figure}
\hrule
\begin{lstlisting}[language=PythonPlus, escapeinside=`']
# self is a cache server
def |integrate|(self, deps):
    # Recursively collect all transitive deps
    all_deps = `$\forall$'[key, vc] that are transitive predecessors of deps (inclusive)
    for k, vcs in all_deps.items():
        # Remove matching versions from I-cache
        consistent_versions =
          self.Inconsistent[k].|remove|(
            filter(vc `$\in$' vcs)
          )
        # Update server's vector clock
        self.vc.|merge_all|(vcs)
        # Merge into C-cache (deps dropped)
        self.Consistent[k].|merge_all|(
          consistent_versions
        )


\end{lstlisting}
\hrule\vspace{2ex}
\caption{Pseudo-code for Dependency integration.}%
\label{fig:alg-pulling}
\end{figure}

After a write is stored in the receiving server's \black{},
  it must eventually become visible on every other server.
This is achieved by two server-side mechanisms: a
  \emph{propagation chain} that disseminates the write to all
  servers,
  and \emph{integration} that promotes the write from
  \black{} to \white{} once it is safe to expose.
Both mechanisms run asynchronously and do not block client
  requests.

\heading{Propagation Chain}
Our data propagation design is inspired by Chain
  Replication~\cite{van2004chain} in distributed systems.
In \sys{}, each chain takes the form
  $S_i \to S_{(i+1)\bmod N} \to \dots \to S_{(i+N-1)\bmod N}
  \to S_i \to \dots \to S_{(i+N-1)\bmod N}$, effectively running
  two full rounds through the $N$ servers.
For example, in a three-server system, there are three chains and
  the chain starting at $S_0$ is
  $S_0 \to S_1 \to S_2 \to S_0 \to S_1 \to S_2$.
Within each chain, a server may function as head, intermediate
  node, or tail, determined by its position in the current chain.

The \texttt{ServerWrite} function in
  Figure~\ref{fig:alg-server} shows how each server processes a
  propagation message.
Writes are propagated in FIFO order along a chain that begins at
  the server handling the initial write and ends at the tail of the
  chain.
As the write moves through the chain, each server adds the write to
  its \black{} and forwards it to the successor during the first
  round; in the second round, the write is simply forwarded.
When the write reaches the tail in the chain, it undergoes
  \textit{integration} (described below), which moves it
  and its dependencies from \black{} into \white{}, making them
  visible to clients.
Steps~\ding{202}--\ding{206} in Figure~\ref{fig:running}
  trace the two-round propagation of $x_1$ from $S_0$ through all
  servers.

The propagation chain implicitly establishes causal relationships
  between writes on different servers: a write can only appear in a
  server's \white{} once all servers in the chain have seen it.
The second round is essential for correctness; we defer the
  detailed discussion to \S\ref{s:cc}.

This propagation takes place asynchronously after the server
  responds to the write, and therefore is not on the critical path.
The tail of the chain has the option to disseminate the write to
  other servers, asking them to integrate the write into their
  \white{}s as well.
The decision whether to inform others or not is a trade-off between
  network cost and visibility and does not affect the correctness
  of the system.

\heading{Integration}
Once a write completes propagation and reaches the tail, it can
  be moved from \black{} into \white{} to become visible.
This is done through \textit{integration}, which merges
  the version along with all of its transitive dependencies into
  \white{}, maintaining the causal cut invariant.
Integration is also triggered when a cache server receives a read
  request: if the client's $deps$ are not yet in \white{}, the
  server integrates them first, ensuring the read is served from
  a causal cut that covers the client's
  dependencies.

The pseudo-code for integration is shown in
  Figure~\ref{fig:alg-pulling}.
Each version only records its nearest (direct) dependencies,
  so integration must recurse: it first integrates the version's
  direct dependencies, then their dependencies, and so on
  (Figure~\ref{fig:alg-pulling} Line 4).
Then, for each dependency set, the integration follows the steps
  below:

\begin{myenumerate2}
\item Iterate over the dependencies in the set (Line 5).
\item For each key-version pair in the dependencies, check if this
  version has already been merged into \white{}.
\item If not, search \black{} for this version, remove it from
  \black{} (Lines 7--10), and merge it into \white{}
  (Lines 14--16) using the same procedure of merging two versions.
  Note that unlike \black{}, \white{} has no dependency metadata;
  the dependencies are automatically dropped when merged into
  \white{}.
\end{myenumerate2}

Integration is a purely local operation: it does not require
  any communication with other servers, and integrated entries are
  removed from \black{}.
The cost is $O(d)$ where $d$ is the size of the transitive
  dependency closure; in practice $d$ is small because most
  dependencies have already been integrated into \white{} by the
  time a version completes propagation, and are skipped during
  the traversal.

For \white{} to remain a causal cut after integration, the
  version itself and all of its transitive dependencies must
  already be present on the same server (in \white{} or \black{}).
This invariant is guaranteed by two-round propagation, as we
  discuss in \S\ref{s:cc} and formally prove in
  \S\ref{subsec:proof}
  (Lemmas~\ref{lemma:deps_are_met}
  and~\ref{lemma:ccache_is_met}).

\subsection{Read Path}\label{s:readpath}

\heading{Client side: issuing the read}
When a client invokes \texttt{Read(key)}, the client library wraps
  it into a \texttt{ClientRead} request that attaches the
  \textit{deps} map as metadata, and sends it to the cache server.

\heading{Server side: serving the read}
When a cache server receives a read request, it first performs
  integration~(\S\ref{s:pull}) on the client's $deps$,
  then returns the value and its version from \white{}, as shown
  in Figure~\ref{fig:alg-server}.
The integration ensures that the client never reads an older
  version than its dependencies.
For example, if a client previously read $y_1$, where
  $x_1 \to y_1$, then when it reads $x$ later, the server
  integrates $y_1$ and recursively $x_1$ into \white{} before
  returning $x$.
This is safe because both $x_1$ and $y_1$ were previously read from some
  server's \white{}, meaning they have already completed propagation
  and are present on every server's \black{} or \white{}.

If the requested key does not exist in either \white{} or
  \black{} (i.e., no client has previously read or written this
  key), the cache server reads it directly
  from the underlying storage and adds the result to \black{} as
  if it were written by a client.
This value will then follow the same propagation chain as a write.

Reads in \sys{} are coordination\hyp{}free: after integration,
  \white{} is a causal cut that covers the client's dependencies,
  so the server can return the value directly without consulting
  other servers.
In Figure~\ref{fig:running}, step~\ding{199} illustrates
  this: $S_0$ integrates $C_2$'s $deps$ ($z_1$) along with all
  transitive dependencies (including $y_1$), and returns $y_1$
  entirely from its local caches, without contacting $S_1$ or
  $S_2$.

\heading{Client side: merging with \textit{local}}
Once the client library receives the server's reply
  $w_{\textit{cached}}$ (a value-version pair), it checks the
  \textit{local} map to see if the same key has been written by
  the client itself.
If not, it returns $w_{\textit{cached}}$ directly.
Otherwise, it merges $w_{\textit{cached}}$ with the version stored
  in \textit{local} and returns the merged result, ensuring that
  the client always observes its own prior writes even if they
  have not yet propagated.
In Figure~\ref{fig:running}, step~\ding{193} shows this in action:
  $C_0$ reads $x$ at $S_2$ before propagation completes, but the
  client library merges the server's response with $x_1$ from
  \textit{local} and returns the correct value.
Finally, the client library adds the server's return version to 
  \textit{deps}.
If the client library fails (e.g., due to a function crash), the
  workflow fails as well; consistency is not violated because no
  stale state is exposed, and a new workflow starts fresh.

\subsection{Read Transactions}\label{s:readtxn}

\begin{figure}
\hrule
\begin{lstlisting}[language=PythonPlus]
# self is a client (workflow)
def |read_txn|(self, keys):
   # Multi-key read with deps for consistency
   values, vcs = |ClientReadTxn|(keys, self.deps)
   res = []
   for k, v, vc in zip(keys, values, vcs):
       # Abort if local conflicts
       if k in self.local \
         and not (self.local[k] <= vc):
           return None  # abort
       res.append(v)
       self.deps[k].|merge|(vc)  # track deps
   return res
\end{lstlisting}
\hrule\vspace{2ex}

\caption{Pseudo-code for read transaction in \sys{}'s client library. The read transaction may abort (Lines 7--10) if keys in the transaction overlap with the client's own writes.}%
\label{fig:alg-client-txn}
\vspace{-8pt}
\end{figure}

\sys{} supports causally-consistent read transactions.
A transactional read request includes a set of keys.
Similarly to single-key reads, read transactions are
  coordination-free: the cache server integrates the client's
  dependencies and reads all keys from its local \white{}, without
  communicating with other servers or waiting for a specific
  version to arrive.
Since \white{} is always a causal cut~(\S\ref{s:dualcache}),
  reading multiple keys from \white{} after integration guarantees
  a causally consistent snapshot.
Figure~\ref{fig:alg-client-txn} presents the pseudo-code for a
  read transaction in the client library.

However, unlike single-key reads, read transactions are not 
    always abort-free.
If the transaction includes a key that the client has previously
  written, the client library merges the server's result with
  \textit{local}, as in single-key reads.
However, since the client's write in \textit{local} may not yet
  have been propagated, the merged value may fall outside the
  causal cut returned by the server.
In this case, the transaction must abort
  (Figure~\ref{fig:alg-client-txn} Lines 7--10).
Aborts are handled by the client library by notifying the scheduler
  and are opaque to the users.
Aborts can only occur in cases where read transactions include keys
  that have been previously written by the same client, such as when
  a client writes $x$ and then reads $x$ and $y$ within a
  transaction.
To prevent aborts, developers can rearrange the order of operations
  by placing writes after read transactions if their keys happen to
  overlap.

\subsection{Causal+ Consistency}\label{s:cc}

This section provides the intuition for why \sys{}
  guarantees CC+.
The formal proof appears in \S\ref{sec:verification}.

The correctness of \sys{} rests on a single invariant: \white{}
  is always a strict causal cut~(\S\ref{s:dualcache}).
Since reads are served from \white{}, this invariant directly
  guarantees that clients observe a causally consistent view.
The invariant is preserved during integration, which moves
  a version together with all of its transitive dependencies from
  \black{} to \white{}.
For this to work, the version and its transitive dependencies
  must already be locally available on all servers (in \black{}
  or \white{}) when integration happens.
If any dependency were missing on some server, that server's
  \white{} would contain a version whose predecessor is not yet
  visible, breaking the causal cut invariant.
Two-round propagation is precisely what ensures this property
  holds, as we explain next.

\begin{figure}
  \centering
  \includegraphics[width=1\columnwidth]{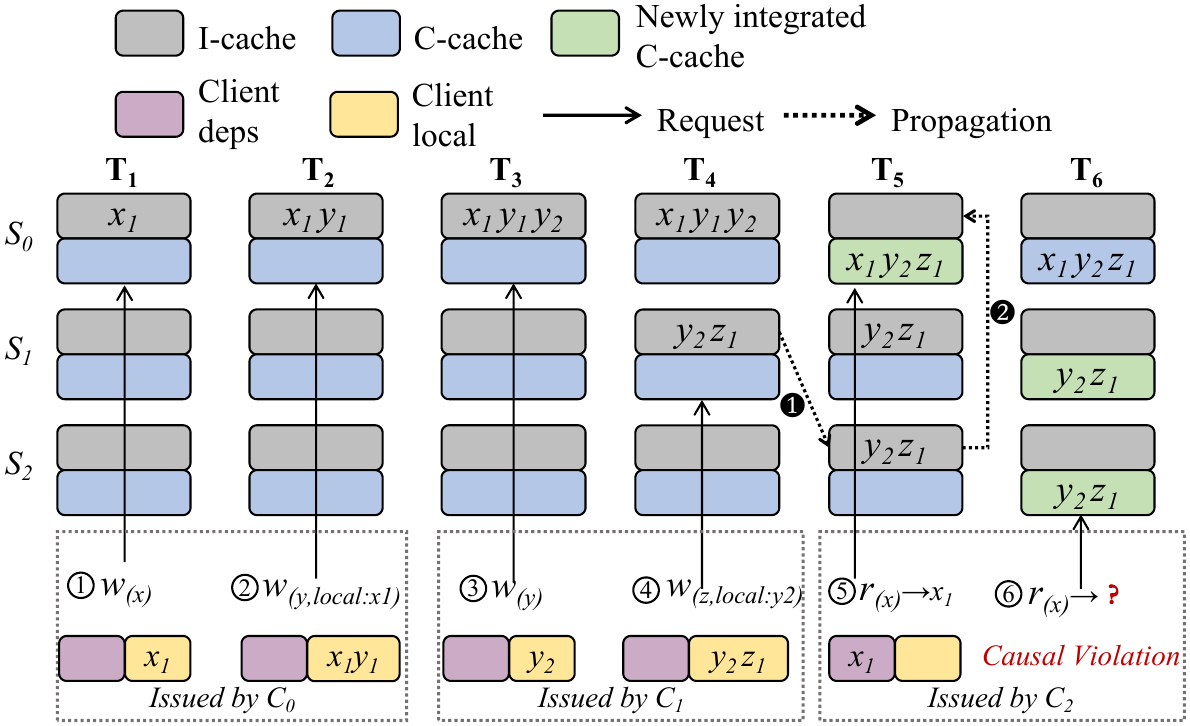}
  \caption{A causal consistency violation under single-round
    propagation.}%
  \label{fig:violation}
\end{figure}

\heading{Single-round propagation is insufficient}
In the original version of \sys{}~\cite{CausalMesh}, a write is
  considered successfully propagated after traversing a chain once.
Upon reaching the tail server, the write is immediately integrated
  into \white{}, along with its dependencies.
During this process, for each key in the dependency set, all
  versions with smaller vector clocks are treated as implicit
  dependencies and integrated as well.
This is necessary because a dependency may record a vector clock
  that results from merging two concurrent versions, and the exact
  version matching that merged clock may not exist; integrating
  all versions with smaller clocks ensures the dependency is
  satisfied.

Figure~\ref{fig:violation} illustrates a concrete violation
  under this design.
Client $C_0$ writes $x_1$ to $S_0$ (step~\ding{192}), then
  writes $y_1$ carrying $x_1$ in \textit{local}
  (step~\ding{193}).
Client $C_1$ writes $y_2$ to $S_0$ (step~\ding{194}), then
  migrates to $S_1$ and writes $z_1$ carrying $y_2$ in
  \textit{local} (step~\ding{195}).
$S_1$ receives $y_2$ via \textit{local} and propagates it along
  with $z_1$ on the chain
  $S_1 \to S_2 \to S_0$.
Under single-round propagation, when $S_0$ receives $y_2$ and
  $z_1$, it immediately integrates them into \white{}.
During this process, integrating $y_2$ causes all earlier
  versions of key $y$ to be integrated as well, including $y_1$
  and its dependency $x_1$.

However, $x_1$ and $y_1$ have not yet been propagated
  to $S_1$ and $S_2$ (e.g., the network connection 
  from $S_0$ to $S_1$ is slow or has failed).
Now client $C_2$ reads $x$ from $S_0$ and observes $x_1$ in
  \white{} (step~\ding{196}).
When $C_2$ subsequently reads $x$ from $S_2$ (step~\ding{197}),
  $S_2$ has not seen $x_1$ and cannot return it.
This is a causal consistency violation: $C_2$ observed $x_1$ on
  one server but cannot observe it in a following read.

\heading{Why two rounds fix the problem}
With two-round propagation, when $S_0$ receives $y_2$ and $z_1$
  at the end of the first round, it does not immediately
  integrate them.
Instead, it continues the second round, forwarding them along
  $S_0 \to S_1 \to S_2$.
Since $x_1$ and $y_1$ were originally written at $S_0$, their
  propagation to $S_1$ was initiated earlier than $z_1$'s second
  round.
By FIFO ordering on the $S_0 \to S_1$ channel, $x_1$ and $y_1$
  arrive at $S_1$ (and subsequently at $S_2$) before $z_1$ does
  in its second round.
Therefore, by the time $z_1$ completes two rounds, $x_1$ has
  already been seen by all servers.
Even if $x_1$ itself has not completed two rounds, its presence
  on all servers ensures it is safe to integrate.

More generally, two rounds guarantee that when a version $v$
  reaches the tail, every version with a smaller vector clock has
  already been observed by all servers, satisfying the critical
  premise above.
The formal proof of this property is given in
  \S\ref{sec:verification}.

Figure~\ref{fig:running} illustrates correct operation under
  two-round propagation.
At step~\ding{199}, $C_2$'s $deps$
  contains $z_1$;
$S_0$ integrates $z_1$ and recursively its dependency $y_1$
  into \white{}, entirely from local caches.
This succeeds because two-round propagation guarantees that
  $y_1$ and $z_1$ have reached $S_0$'s \black{} by the time
  $C_2$'s read arrives.

\section{\sys{}-TCC}\label{s:tcc}

\sys{} only supports read transactions within a single serverless function.
However, certain workloads necessitate read-write transactions,
  specifically when dealing with access-control lists (ACLs).
For example, a workflow that updates a document and
then tightens the ACL protecting it must commit both writes
atomically (or must update the ACL before updating the document).
Without atomicity, a soon-to-be-revoked user can observe the new document under the old ACL.
To tackle this limitation, we propose an extension of \sys{} called \sys{}-TCC, 
  which provides \emph{Transactional Causal Consistency} (TCC)~\cite{akkoorath2016cure,lloyd2011cops} as a stronger 
  consistency level.
In \sys{}-TCC, the whole workflow is treated as a transaction.

\begin{figure}
\hrule
\begin{lstlisting}[language=PythonPlus, escapeinside=`']
# self is a cache server
def |integrate_tcc|(self, deps):
    # Collect transitive deps
    all_deps = `$\forall$'[key, vc] that are transitive predecessors of deps (inclusive)
    for k, vcs in all_deps.items():
        # Remove from I-cache
        consistent_versions =
          self.Inconsistent[k].|remove|(
            filter(vc `$\in$' vcs)
          )
        self.vc.|merge_all|(vcs)  # update vc
        # Append new C-cache version
        new_version =                                 self.Consistent[k].|merge_all|(
           consistent_versions
          )
        self.Consistent[k].append(new_version)

def |ClientReadTCC|(self, key, deps):
   self.|integrate_tcc|(deps)  # integrate deps
   # Find a consistent version
   for v in self.Consistent[key]:
       if deps \/ v \/ v.deps is Cut:
           return (v.value, v.vc)
   return None  # no consistent version
\end{lstlisting}
\hrule\vspace{3ex}
\caption{Pseudo-code for \sys{}-TCC's Server.}%
\label{fig:alg-tcc}
\vspace{-8pt}
\end{figure}

TCC and CC+ differ in two key aspects. 
First, TCC ensures atomicity of writes, meaning that all writes from a 
  transaction are either fully visible or not visible at all. 
In contrast, CC+ does not offer such atomicity guarantees. 
Second, TCC enforces that all reads within a transaction must originate from 
  the same causal cut. 
For example, if a client reads $x=0$, all subsequent reads of $x$ within the 
  same transaction will also return $0$. 
On the other hand, CC+ allows for the possibility of reading newer values in
  subsequent read operations by reading from a monotonic cut.
In practice, CC+ is sufficient for many applications,
  including the movie review service in
  \S\ref{s:eval:real}.
Developers should use \sys{}-TCC when their workflows require
  multi-key atomic writes or cross-function transactional
  guarantees, at the cost of potential aborts.

To enforce atomic writes, \sys{}-TCC's client library saves writes
  in a buffer and returns to the client immediately.
The writes are then sent to the server in a batch at the end of the
  workflow.
If the workflow has multiple leaves, it will add a dummy sink
  function that joins all leaves.
The entire batch shares a single vector clock and a single
  dependency set (the accumulated $deps$ and $local$ of the
  workflow), and is propagated along the chain as one unit.
During integration, the cache server integrates all writes in the
  batch atomically: either all of them are moved into \white{},
  or none are.
This ensures that other clients never observe a partial subset of
  a transaction's writes.
Because the batch is treated as a single propagation unit, write
  batching does not introduce additional complexity to causal
  dependency tracking or vector clock management compared to
  single-key writes.

To make all reads come from the same cut, \sys{}-TCC extends \white{} to 
  be a map from a key to a list of tuples that includes the value, VC, and deps.
\[
    \begin{array}{rcl}
        \textit{\white{}} &:=& \{Key \mapsto [~(Value, VC, Deps)~] \}
    \end{array}
\]
\begin{figure*}[tb]
{%
\resizebox{\textwidth}{!}
{
\small
\begin{tabular}{lcccccc}
  & \textbf{Consistency} & \textbf{Unk. ReadSet} & \textbf{Coordination Cost} & \textbf{Read / Write} & \textbf{Abort Free} & \textbf{Visibility} \\
\toprule
\sys{} & CC+ & Yes & 0 RTT & 0 RTT / 1 RTT to DB & Yes &  $2N$ RTT \\
\sys{}-TCC & TCC & Yes & 0 RTT & 0 RTT / 1 RTT to DB & No & $2N$ RTT \\
HydroCache-Con & TCC & No & 2 RTTs & 0 RTT / 1 RTT to DB & Yes & refresh period \\
HydroCache-Opt & TCC & No$^*$ & 0 RTT $\sim 2N$ RTT & 0 RTT / 1 RTT to DB & No & refresh period \\
FaaSTCC & TCC & Yes & 0 RTT $\sim 2N$ RTT & 0 RTT / 1 RTT to DB & No & refresh period \\
\bottomrule
\end{tabular}
}
}
\vspace{1ex}
\caption{Comparison between \sys{}, \sys{}-TCC, HydroCache-Con, and
HydroCache-Opt. $N$ is the number of servers. Unknown ReadSet means that the read set does not need to be known ahead of time, which is needed for supporting dynamic workflows. HydroCache-Opt's Unknown
ReadSet field is No$^*$ because it supports partially dynamic workflows (\S\ref{s:discuss}). 
In HydroCache and FaaSTCC, writes become visible after a refresh period, set to 100ms and 50ms in the original papers.}%
\label{fig:table:compare}
\end{figure*}

Figure~\ref{fig:alg-tcc} shows \sys{}-TCC's pseudo-code on the server side.
During dependency integration, rather than updating the value in \white{} in
  place as \sys{} does, the cache server in \sys{}-TCC creates a new version
  and appends it to the list so that the list contains multiple versions for
  each key (Figure~\ref{fig:alg-tcc} Lines 13--16).
Retaining historical versions allows subsequent reads in an
  ongoing transaction to find a version consistent with what the
  transaction has already observed.
Upon receiving a read request, the cache server searches the
  version list for a version that is consistent with all versions
  the workflow has previously read (i.e., a version such that the
  union of all read versions still forms a causal cut, as shown in Figure~\ref{fig:alg-tcc}, Line 22).
If such a version exists, it is returned; otherwise, the workflow
  must abort and retry.
In the case of workflows with multiple parallel functions,
  \sys{}-TCC runs a validation phase at the sink function (where
  the parallel branches join): it checks whether the union of all branches' read sets forms a causal cut.
If it does not, the workflow is aborted and retried.
All aborts are handled by the client library and are opaque to  developers.

To prevent the version list from growing unboundedly, we
  implement it as a ring buffer of fixed size.
When the buffer is full, the oldest version is evicted.
The buffer size is a tunable parameter that controls the trade-off
  between abort rate and memory usage: a larger buffer retains more
  historical versions, reducing the chance that a consistent
  version has been evicted, while a smaller buffer saves memory at
  the cost of higher abort rates.

\section{Implementation}\label{s:impl}

We implemented \sys, \sys-TCC, and two baselines.

\heading{Baselines: HydroCache and FaaSTCC}
HydroCac\hyp{}he\cite{wu2020hydrocache} is currently the state-of-the-art 
  serverless caching system.
It guarantees Transactional Causal Consistency (TCC). 
It has two versions: a conservative version (HydroCache\hyp{}Con) and an optimistic 
  version (HydroCache-Opt). 
In HydroCache-Con, prior to execution, a centralized scheduler distributes the read set to all candidate cache servers and blocks until it has received their responses with their respective snapshots of the read set. The scheduler then uses these responses to construct a consistent causal cut and send it back to all cache servers.
In HydroCache-Opt, the scheduler checks for causal violations between two 
  serverless functions during execution. 
If a violation is detected, the entire workflow is aborted and retried.
FaaSTCC improves HydroCache-Opt by replacing dependency metadata with a \textit{snapshot interval} which is a time frame where reads are valid, however, it requires the underlying storage system to provide a global timestamp.
Figure~\ref{fig:table:compare} shows the full comparison between them.

\heading{Prototype}
We implement \sys{}'s cache server in Rust ($\sim$5K lines) and
  its client library in Go ($\sim$400 lines).
The two components do not share any in-process code; they
  communicate exclusively via gRPC~\cite{grpc}, which is
  language-agnostic, so no cross-language overhead is introduced.
As HydroCache is not open-source, we also implement it in Rust
  and ensure that it achieves the same or better performance as
  reported in the original paper~\cite{wu2020hydrocache}.

HydroCache and FaaSTCC have a background thread that uses double-buffered hash tables~\cite{evmap} 
  to refresh the cache: the background thread updates one table while read 
  handlers in the main thread read the other; an atomic pointer swap exposes new writes
  ensuring that the refresh does not affect readers in the critical path.

\section{Evaluation}\label{s:eval}
\sys{} helps serverless developers rely on caches without the complexity of weak consistency semantics.
To see how well \sys{} works, we answer the following questions:

\begin{myitemize2}
    \item What is \sys{}'s performance on micro-benchmarks and how does it compare to prior serverless cache systems?~(\S\ref{s:eval:micro})
    \item What overhead does \sys{} introduce?~(\S\ref{s:eval:overhead})
    \item How does \sys{}-TCC perform with varying data contention and buffer configurations?~(\S\ref{s:eval:contention})
    \item How does \sys{} scale with server count?~(\S\ref{s:eval:scale2})
    \item What are the latency and throughput of representative applications running on \sys{}?~(\S\ref{s:eval:real})
    \item How long does it take for a write in \sys{} to become visible on other servers?~(\S\ref{s:eval:visibility})
\end{myitemize2}

\subsection{Experimental Setup}
In our evaluation, we used CloudLab~\cite{cloudlab} m510 machines with 8-core 2.0 GHz CPUs, 64GB of RAM, 256GB NVMe SSDs, and 10\,Gbps NICs. The typical round-trip time (RTT) between servers is 0.15ms. We use Nightcore~\cite{jia2021nightcore} as the serverless runtime. 
It uses two cores and 8 workers per machine for all experiments except for the experiments that evaluate 
  the real-world applications (\S\ref{s:eval:real}) which use five cores and 16 workers.
We run a Redis~\cite{redis} server in \textit{append-only} mode as our 
  underlying storage, with a custom conflict resolution layer on top of it. We add an artificial latency of 5ms to the Redis server using \texttt{netem} to emulate remote storage so that it has similar latency to those found in public clouds (e.g. AWS Lambda with DynamoDB). For all evaluations except the scalability evaluation, we used a setup with three workers, consisting of five machines: one machine running Redis as the database, one machine running a client and a scheduler, and three machines each running a Nightcore instance and a cache (either \sys{} or HydroCache).

\sys{} uses a single thread. The ring buffer size in \sys{}-TCC is set to one to have a comparable memory footprint. Both versions of HydroCache and FaaSTCC use an additional thread to run a background task that refreshes the cache by merging new updates from the database. The refresh period was set to 100ms and 50ms, respectively, as in the original papers. Unless otherwise specified, the caches were pre-warmed to remove the overhead of data retrieval from the persistent database.

We use wrk2~\cite{wrk2}, which is a constant-load
  HTTP workload generator and measurement tool
  to obtain the latency and throughput numbers.
Each workload runs for a total of 90 seconds, with the first 30 seconds serving as a warm-up period. The results of the subsequent 60 seconds are reported.

Compared to our prior version of 
CausalMesh~\cite{CausalMesh}, the
  protocol now uses two-round propagation
  (\S\ref{s:propagation}).
Since the second round is asynchronous and off the critical path,
  it does not affect read/write latency or throughput; its primary
  impact is on the visibility window, which we evaluate in \S\ref{s:eval:visibility}.

\subsection{Micro-Benchmark}\label{s:eval:micro}
In our micro-benchmark, we evaluate a three-function serverless workflow that aligns with the one described in the HydroCache paper~\cite{wu2020hydrocache}. The first two functions in the workflow read three keys, while the last function writes to a single key. The keys are sampled from a pool of 1,000,000 keys, following a Zipfian distribution with a coefficient of 1.0. The value is an 8-byte string.

\begin{figure}
  \centering
  \includegraphics[width=\columnwidth]{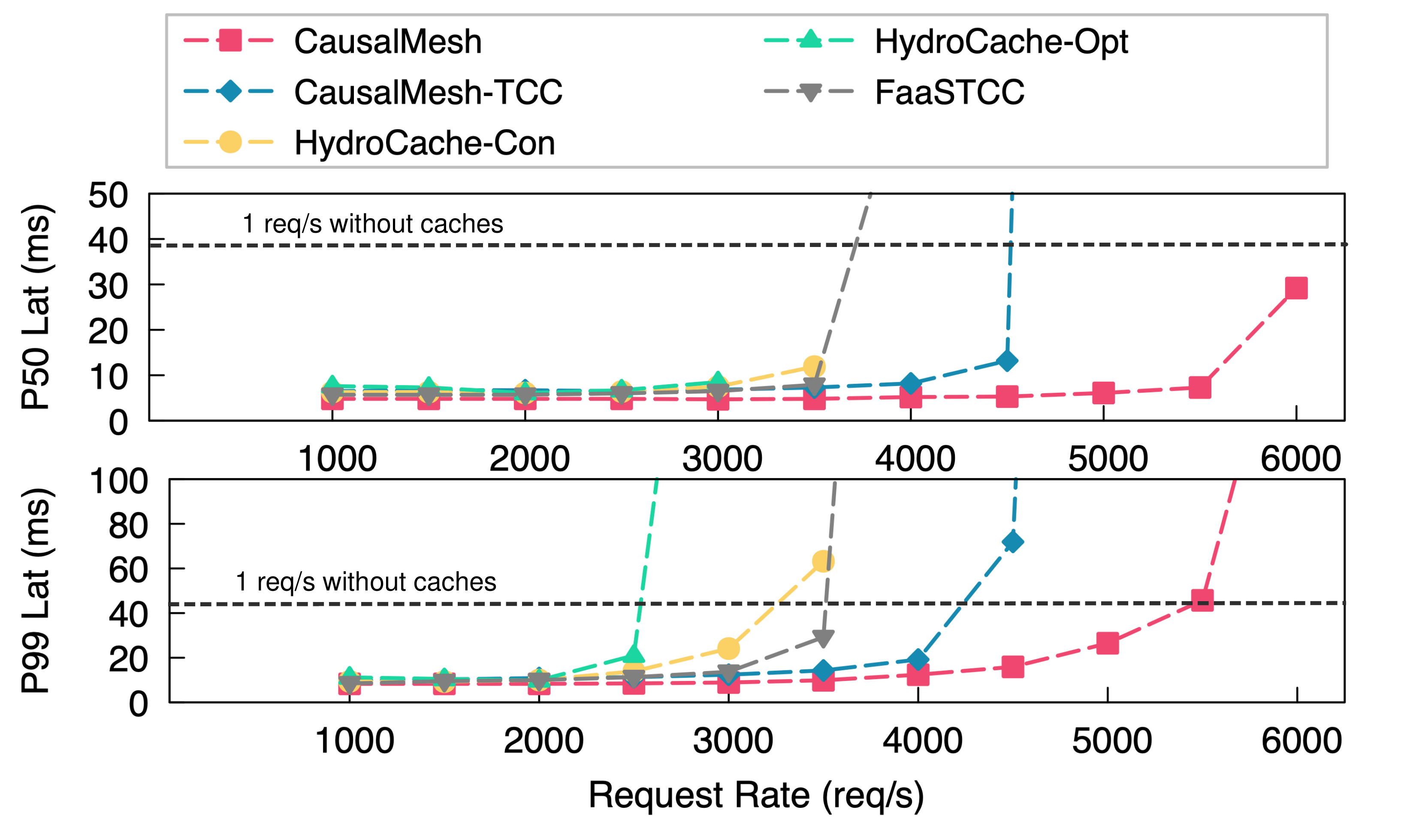}
  \caption{Median and tail response time and throughput of \sys{}, \sys{}-TCC, HydroCache-Con, and HydroCache-Opt in the micro-benchmark.}%
  \label{fig:short}
\end{figure}

\heading{Results}
Figure~\ref{fig:short} shows the results of our micro-benchmark. 
Compared to HydroCache and FaaSTCC, \sys's throughput is 1.57$\times$--2.2$\times$ higher.
In terms of median latency, \sys{}'s is 7\%--59\% lower than HydroCache-Con,
  15\%--44\% lower than HydroCache-Opt and 5\%--38\% lower than FaaSTCC. 
Regarding tail latency, \sys{} achieves up to 85\%, 97\% and 54\% lower latency
  than HydroCa\-che-Con, HydroCache-Opt and FaaSTCC, respectively.
\sys{}-TCC achieves comparable latency to HydroCache and FaaSTCC
  but much better throughput: up to 1.35$\times$ and 1.6$\times$ higher compared to HydroCache and FaaSTCC.
For comparison, Figure~\ref{fig:short} also shows a lower bound on the latency of the workflow when accessing the database directly without caches (horizontal dotted line).

\heading{Takeaway}
Caches are critical in keeping the latency of stateful serverless functions low (up to 4$\times$ lower than the baseline
without caches shown as a dotted horizontal bar).
\sys{}-TCC provides the same consistency guarantees and supports the
  same applications as state-of-the-art serverless caches but achieves considerably higher throughput. 
\sys{} further achieves lower latency at the tail if the workflow
  does not require read/write transactions within a function or 
  cross-function transactions.

\subsection{Resource Overhead}\label{s:eval:overhead}
To quantify the cost of running \sys, we send requests to the serverless 
  workflow at a constant rate of 
  1000 requests/second, using the same workload as that in the micro-benchmark, and we analyze the overhead from two sources: CPU and memory usage.
For memory, we differentiate between the total usage of the cache server and the size of internal metadata (i.e., how much additional data is required to ensure correctness).
In \sys{}, metadata includes the contents of the \black{}, while in \sys{}-TCC, it includes the \black{} and 
any dependencies in the \white{}.
In FaaSTCC, it includes the interval timestamps.
In HydroCache, it includes all dependencies of any element in the cache.
Note that, absent a separate garbage collection protocol, HydroCache's metadata will grow infinitely (\S\ref{s:discuss});
we measure the size of HydroCache's metadata after 1 minute.

\begin{figure}
{
\resizebox{\columnwidth}{!}{
\begin{tabular}{lccc}
\textbf{} & \textbf{\sys{} / -TCC} & \textbf{HydroCache-Con/Opt} & \textbf{FaaSTCC} \\
\toprule
\textbf{CPU (\%)}  & \textbf{48.4} / 60.1 & 98.3 / 54.4 & 52.9 \\
\textbf{Memory (MB)}  & \textbf{47.9} / 50.3 & 69.3 / 69.6 &  61.0 \\
\textbf{Metadata (KB)}  & 35.1 / 69.9 & 45.3 / 99.8 & \textbf{28.8} \\
\bottomrule
\end{tabular}
}
}
\vspace{1ex}
\caption{CPU and memory usage of cache servers in the micro-benchmark. The request rate is 1000 requests per second. Our system uses a single thread, while HydroCache and FaaSTCC use two threads. Metadata is the additional data required by each protocol to ensure correctness.}
\label{fig:table:cpu}
\end{figure}

\heading{Result}
Figure~\ref{fig:table:cpu} shows that \sys{} has 50.6\% lower CPU consumption compared to HydroCache-Con, 11\% lower compared to HydroCache-Opt and 8.5\% lower compared to FaaSTCC. For \sys{}-TCC, the CPU consumption is 39\% lower than HydroCache-Con; it is 10\% higher than HydroCache-Opt and FaaSTCC. In terms of memory, \sys{} consumes up to 30\% less memory than HydroCache and FaaSTCC, while \sys{}-TCC uses up to 27\% less memory. The sizes of metadata in \sys{} and \sys{}-TCC are 2.2\% and 4.5\% of the total data set, respectively.

\heading{Takeaway}
\sys{}(-TCC) incurs less CPU and memory overhead compared to HydroCache and FaaSTCC.
The metadata of \sys and FaaSTCC stays low and stable while HydroCache's metadata grows over time. FaaSTCC reduces a great amount of metadata by delegating the job of assigning versions to the underlying storage system.

\subsection{Contention and Abort Rate}\label{s:eval:contention}
We evaluate \sys{}-TCC's performance under contention at a constant
  load of 4,000 requests/second, the system's saturation point.
We vary the Zipfian skew (1.0, 1.25, 1.5), which controls the
  contention, and the ring buffer size (1, 4, 16) to explore the
  trade-off between memory usage and transaction success rate.
The ring buffer size is the number of recent versions \white{}
  retains per key (\S\ref{s:tcc}); a larger buffer increases
  the likelihood that a transaction finds a version consistent with
  its prior reads, reducing aborts.

\begin{figure}
{
\resizebox{\columnwidth}{!}{
\begin{tabular}{l ccc}
& \textbf{Buffer Size 1} & \textbf{Buffer Size 4} & \textbf{Buffer Size 16} \\
\toprule
\textbf{Zipf 1.0}  & 6.5  & 1.8  & 1.4  \\
\textbf{Zipf 1.25} & 22.0 & 9.3  & 8.3  \\
\textbf{Zipf 1.5}  & 65.5 & 21.6 & 17.5 \\
\bottomrule
\end{tabular}
}
}
\vspace{1ex}
\caption{\sys{}-TCC's abort rates (\%) across varying Zipfian skewness factors and buffer sizes.}
\label{fig:table:contention}
\end{figure}

\heading{Result}
Figure~\ref{fig:table:contention} shows that increasing the buffer
  size consistently lowers abort rates across all skew levels.
At Zipf 1.0, the abort rate drops from 6.5\% to 1.4\% as the buffer
  grows from 1 to 16 slots.
This effect scales with contention: under high skew (Zipf 1.5), a
  16-slot buffer slashes the abort rate from 65.5\% to 17.5\%, a
  3.7$\times$ improvement.

\heading{Takeaway}
Most gains occur when moving from a buffer size of 1 to 4, suggesting
  that even a small buffer effectively captures the majority of
  immediate conflicts.

\subsection{Effect of the Number of Caches}
\label{s:eval:scale2}
To evaluate how \sys{} scales with the number of servers, we
conduct experiments with 2--16 servers (the same order of magnitude as AWS DAX's maximum of 11 cache nodes).
More servers result in more concurrent clients.
We issue requests in increments of 50 req/s until the system is nearly saturated, 
  which we determine by observing a tail latency longer than 10ms. 
Each function randomly reads from two keys and writes to one key.

\begin{figure}[t]
  \centering
  \includegraphics[trim=0.1cm 0.8cm 0.4cm 1.2cm, clip, width=\columnwidth]{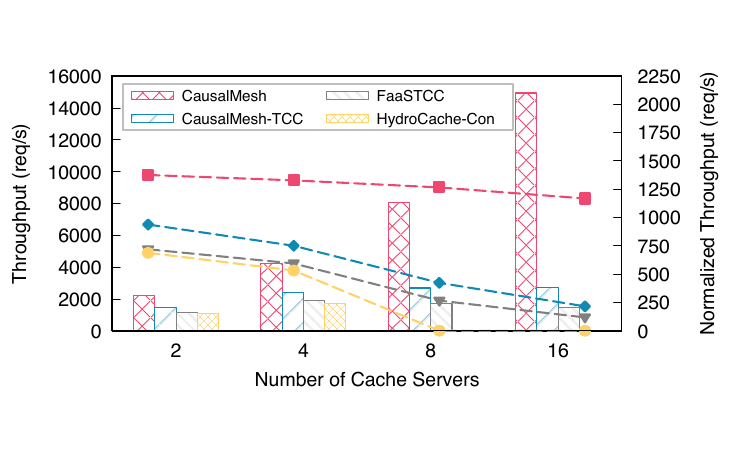}
  \caption{The histogram with y-axis on the left depicts the throughput as we vary the number of servers. The line plot with y-axis on the right shows the normalized throughput by dividing the throughput by the number of servers. }%
  \label{fig:scale2}
\end{figure}

\heading{Results}
We normalize the results by dividing the raw throughput by the number of
servers. Figure~\ref{fig:scale2} includes a histogram illustrating the raw
throughput and a line plot depicting the normalized throughput. It shows that
\sys{}'s normalized throughput is nearly constant, which means \sys{} scales
almost linearly with respect to the number of servers. On the other hand,
\sys{}-TCC reaches saturation at around 2800 requests/second due to increased contention. 
FaaSTCC experiences throughput degradation as the number of servers increases. 
\sys{}-TCC achieves 1.3$\times$--1.8$\times$ higher throughput than FaaSTCC. 
Both HydroCache-Opt and HydroCache-Con perform worse than both FaaSTCC and \sys{}-TCC; HydroCache does not scale to 8 servers or beyond because 
the cost of coordinating between those servers and pulling dependencies is far too high. HydroCache-Opt performs even worse, which is
why we do not include it in the figure.

\heading{Takeaway}
Developers should use \sys{} whenever allowed, as it has better performance when there are more cache servers;
developers should only use \sys{}-TCC when read transactions across
multiple serverless functions or read-write transactions are necessary. We discuss scalability further in \S\ref{s:discuss:scale}.

\subsection{Movie Review Service}\label{s:eval:real}
We evaluate \sys{}'s performance on the 
  movie review service described in DeathStarBench~\cite{gan19open,deathstarbench}.
In this service, users create accounts, read reviews, view the plot and 
  cast of movies, and write movie reviews. 
We use Beldi's implementation of this app~\cite{Beldi} which is a V-shape workflow of 13 serverless functions.

We evaluate a mixed workload, consisting of 50\% ComposeReview and 50\% ReadReview. 
ComposeReview generates a review for a random user and movie, and then saves the 
  review ID to the profiles of both the movie and the user. 
ReadReview involves two functions.
First, it reads the profile of a movie to retrieve all associated review IDs. 
Then, it reads the contents of the reviews using those IDs. 
It is worth noting that HydroCache-Con cannot support this type of workload as 
  it requires prior knowledge of the keys. 

\begin{figure}
  \centering
  \includegraphics[width=\columnwidth]{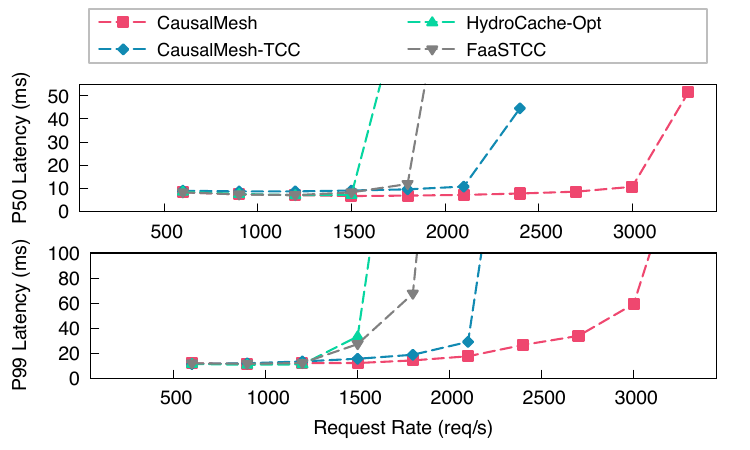}
  \caption{Comparison of \sys{}, HydroCache, and FaaSTCC in terms of median and tail response time and throughput in a mixed workload that has contention between reads and writes.}%
  \label{fig:media2}
\end{figure}

\heading{Results}
Figure~\ref{fig:media2} shows that both HydroCache-Opt and FaaSTCC start experiencing high tail latency 
  at around 1,500 req/s. 
In contrast, \sys{} achieves 2$\times$ higher throughput while reducing median 
  latency by up to 10\% and tail latency by up to 64\% before HydroCache-Opt and FaaSTCC
  become saturated. 
\sys{}-TCC achieves up to 1.35$\times$ higher throughput and similar latency.

\heading{Takeaway}
Both \sys{} and \sys{}-TCC outperform HydroCache and FaaSTCC in throughput for real-world
  applications. 
As Causal+ consistency is sufficient for many applications, including the 
  movie review service above, \sys{} significantly 
  reduces the latency when compared to the others.

\subsection{Visibility}\label{s:eval:visibility}
\begin{figure}
\centering
{
{
\small
\begin{tabular}{lcccccc}
\textbf{Num Servers} & 3 & 4 & 5 & 6 & 7 & 8 \\
\toprule
Inconsistency & \multirow{2}{*}{2352} & \multirow{2}{*}{2922} & \multirow{2}{*}{3497} & \multirow{2}{*}{4115} & \multirow{2}{*}{4662} & \multirow{2}{*}{5290} \\
Window ($\mu$s) & & & & & &\\ 
\midrule
Marginal ($\mu$s) & - & 570 & 575 & 618 & 547 & 628 \\ 
\bottomrule
\end{tabular}
}
}
\vspace{1ex}
\caption{Relationship between the number of servers and visibility, as measured by the inconsistency window~\cite{bermbach2011eventual}. Marginal inconsistencies indicate the increase in delay that the system experiences when an additional server is added.}
\label{fig:table:vis}
\end{figure}

To evaluate the visibility of \sys{}, we use the concept of observed inconsistency window~\cite{bermbach2011eventual}. We compute the inconsistency window using the following steps:
\begin{myenumerate2}
\item Create a timestamp, $t_1$.
\item Write to a server and save the version received from it.
\item Create another timestamp, $t_2$.
\item Poll the result from the tail server until it sees the saved version.
\item Log the elapsed time from the mean of $t_1$ and $t_2$.
\end{myenumerate2}
We vary the number of servers and record the effect this has on the inconsistency window. Additionally, we calculate the marginal inconsistency window by subtracting the inconsistency window of $N$ servers from that of $N-1$ servers.

\heading{Results}
Figure~\ref{fig:table:vis} shows that the marginal inconsistency
  window remains stable between 500 and 650~$\mu$s, indicating that the
  visibility is nearly proportional to the number of servers in the
  system.
  The additional round of chain traversal required for
  correctness~(\S\ref{s:cc}) nearly doubles the inconsistency
  window: at the 3-server default configuration, it grows from
  approximately 1.3\,ms in the single-round conference
  version~\cite{CausalMesh} to 2.4\,ms in the current two-round
  design.

\heading{Takeaway}
In \sys{} and \sys-TCC, the inconsistency window grows as the number of servers
  increases.
\sys{} exhibits a significantly lower inconsistency window
  (e.g., 5ms in an eight-server configuration).
In contrast, HydroCache, FaaSTCC, and other systems with background refreshing can have 
  an inconsistency window up to the refresh period, which is 100ms in HydroCache and 50ms in FaaSTCC. 
Long refresh periods allow them to trade performance by
  tolerating increased staleness.
Lowering their refresh period to match \sys{}'s
  single-digit millisecond-scale inconsistency window would cause them to be unresponsive, as frequent polling
  creates prohibitive resource contention with write-serving
  threads.
\sys{}'s push-based propagation avoids this trade-off.

\subsection{How CausalMesh-TCC Outperforms Others}
CausalMesh-TCC's performance advantages stem from several key factors. 
Figure~\ref{fig:table:compare} lists the characteristics of CausalMesh, HydroCache, and FaaSTCC.
One visible advantage of CausalMesh-TCC is the absence of coordination costs, along with better data freshness.
This is a crucial aspect, as delays in data visibility increase the likelihood of missing versions when a client migrates to a new server.
Another significant factor is that CausalMesh eliminates the need for a background thread that periodically updates data, a feature present in both HydroCache and FaaSTCC.
The background thread is used to subscribe to the underlying database and apply new writes to the cache data structure.
The new writes must be applied atomically or in a causal order to maintain correctness, thus creating contention with the request-serving threads of a cache server.
We managed to considerably reduce this contention by implementing a double-buffered hash table (\S\ref{s:impl}) that improved upon the published designs.
However, despite our efforts, when the request rate is high, our optimized versions of prior work are still hamstrung by significant overheads.

\section{Formal Verification}\label{sec:verification}
This section presents the formal verification of \sys{}'s
  correctness.
We first describe how the formal verification effort led to the
  discovery of a critical bug in the original protocol and
  motivated the two-round propagation redesign
  (\S\ref{subsec:violation}).
We then present our TLA-style modeling approach
  (\S\ref{subsec:modeling}), followed by the proof of correctness
  organized around three key proof steps
  (\S\ref{subsec:proof}).
Finally, we discuss the Dafny verification effort, including the
  key challenges encountered (\S\ref{subsec:dafny_verification}).

\subsection{Bug Discovery}\label{subsec:violation}

In the original version of \sys{}~\cite{CausalMesh}, a write
  was considered successfully propagated after traversing a chain
  once.
As described in \S\ref{s:cc} (see Figure~\ref{fig:violation}),
  this single-round design introduces a subtle causal consistency
  violation, where a version is integrated at the tail before its
  transitive dependencies reach all servers, causing a migrating
  client to observe inconsistent state.
We addressed this by redesigning the protocol to use two rounds
  of propagation~(\S\ref{s:propagation}).

We discovered this violation during the Dafny proof that
  every version in \white{} is \globalmet.
Under the single-round design, we needed to show that when a
  version $v$ reaches the tail and triggers integration, all of
  $v$'s transitive dependencies are already present on every server.
This property cannot be established, because a dependency $u$ of
  $v$ originating from a different server may not have completed
  its own chain traversal by the time $v$ reaches the tail.
After several alternative auxiliary invariants all failed at the
  same proof obligation, we concluded that the issue lies in the
  protocol itself: single-round propagation does not guarantee
  that transitive dependencies are globally available at
  integration time.
This led us to redesign the protocol with two-round propagation,
  after which the proof went through.

This violation went undetected despite the original design
  being specified in TLA+~\cite{tla+} and verified with the TLC
  model checker~\cite{TLC}, because bounded model checking
  explores only a finite state space (e.g., small node counts and
  limited interleaving depths), and the absence of a
  counterexample within those bounds does not imply correctness
  beyond them.
In contrast, deductive verification in Dafny failed at a precise
  proof obligation that pinpointed the flawed protocol
  assumption, illustrating how a stuck proof in deductive
  verification can serve as a diagnostic tool for protocol
  design.

\subsection{Modeling \sys{}}\label{subsec:modeling}

Distributed systems are naturally modeled as state machines,
  where the system evolves through transitions triggered by events
  such as message delivery or internal processing.
This approach is well suited for verification because it supports
  \emph{inductive reasoning}: one can prove a property holds in all
  reachable states by showing it holds initially and is preserved
  by every transition, without enumerating executions.
We specify \sys{} in the Temporal Logic of Actions
  (TLA)~\cite{tla} style with Dafny~\cite{leino2010dafny} as the
  host language, as in prior verified
  systems~\cite{ironfleet,automan}.
The specification extends our previous TLA+ model with
  two-round propagation (\S\ref{s:writepath}).

In the resulting Dafny specification, each \emph{state} captures a
  snapshot of the system, including all cache servers' and clients'
  local states, as well as a global \emph{environment} that records
  the set of messages among nodes.
Formally, a state $s$ is a collection of the following:
\[
s = \{ \mathsf{ServerStates},\ \mathsf{ClientStates},\
  \mathsf{Environment} \}.
\]

\begin{figure}
\hrule
\begin{lstlisting}[style=dafnystyle]
# s is the old state of a cache server
# s' is the new state of a cache server
# m is the received message
# sm is the sent message
predicate ClientRead(s, m, s', sm)
    requires m is Read
{
    var (icache', ccache') :=                    integrate(s.icache, s.ccache, m.deps);
    && s' == s.(
        icache := icache', 
        ccache := ccache'
       )
    && sm == Read_Reply(s'.ccache[m.key])
}
\end{lstlisting}
\hrule\vspace{3ex}
\caption{The TLA-style action predicate for the ClientRead function in Figure~\ref{fig:alg-server}.}
\label{fig:action}
\end{figure}

The system evolves through a set of \emph{actions}, each of which
  captures the atomic behavior of a single node and specifies how
  the system state is updated in one step.
We define an action for every event a node can process: a client
  issuing a read or write, and a server processing a read, write,
  or propagation message.
For example, Figure~\ref{fig:action} shows the action for a cache
  server processing a read request, which corresponds to the
  \texttt{ClientRead} function in Figure~\ref{fig:alg-server}: on
  receiving a read message, the server integrates the
  dependencies, updates its caches, and returns a read reply.

Based on node-level actions, the system-level transition is defined
  as the disjunction of all enabled node-level actions.
That is, in each step, one node performs an action, resulting in a
  new global system state.
A system \emph{behavior} is thus an infinite sequence of such
  transitions.
As outlined above, we prove correctness by identifying key
  invariants and showing they are preserved by every action,
  establishing that all possible execution traces satisfy causal
  consistency.

\subsection{Proof of Correctness}\label{subsec:proof}

\heading{Proof overview}
Our goal is to prove that any client reading from any server always
  observes a causally consistent view.
The proof proceeds in three steps:

\begin{myenumerate2}
\item \textbf{Propagation safety
(Lemmas~\ref{lemma:propagated}--\ref{lemma:propagates_implies_all_previous_versions_met}).}
Two-round propagation guarantees that when a version completes
propagation, it and all versions with smaller vector clocks
(including its transitive dependencies) have been observed by
every server.
This is what makes coordination-free integration
possible.

\item \textbf{Integration safety
(Lemmas~\ref{lemma:versions_before_pvc_are_met}--\ref{lemma:ccache_is_causal_cut}).}
Whenever integration is triggered, whether by a completed
propagation or a client read, the dependencies to be integrated
are \globalmet{} and can be found locally.
As a result, \white{} is always a causal cut after integration.

\item \textbf{Client-side guarantees
(Theorems~1--4).}
Since \white{} is a causal cut and the client library merges
server replies with its own $local$ writes, the four CC+
properties (Read Your Writes, Monotonic Read, Writes Follow Reads,
Monotonic Writes) follow directly.
\end{myenumerate2}

All lemmas are proved by \emph{inductive invariant preservation}:
  we show each property holds in the initial state and is preserved
  by every possible system action.
Since we do not fix the number of servers nor the number of clients, the proof establishes correctness for \emph{all} system
  configurations. This is the advantage over bounded model checking, which can only verify finite instances.

\heading{Notation}
Let $C$ denote a client and $S$ a cache server.
Each client $C$ maintains two local structures: a dependency map
  $deps$, which maps each previously read key to the vector clock
  ($vc$) of its latest observed version, and a local write buffer
  $local$, which maps each previously written key to a list of
  written versions, where each version is a tuple consisting of the
  value ($val$) and the vector clock
  ($vc$)~(\S\ref{s:clientlib}).
Each server $S$ maintains a \textit{\white{}} and a \textit{\black{}}
  as we introduced in \S\ref{s:dualcache}.

\begin{definition}[\Globalmet]
We say that a version $v$ is \emph{\globalmet{}} if, for every
  server $S$, either $S.\textit{\white{}}$ contains a version of
  $v.key$ with vector clock $vc' \geq v.vc$, or $v$ exists in
  $S.\textit{\black{}}$.
\end{definition}

\begin{definition}[Propagation Path]
A \emph{propagation path} is a sequence of servers
  \((S_{0}, S_{1}, \ldots, S_{n})\) that a version has been
  propagated to.
For every adjacent pair in the sequence, the version is propagated
  from server \(S_i\) to server \(S_{i+1}\), and \(S_{i+1}\) is
  the successor of \(S_{i}\).
\end{definition}

\begin{definition}[Complete Chain]
A \emph{complete chain} is a propagation path that contains all
  cache servers in the system, starting from some server $S$ and
  ending at $S$'s predecessor.
\end{definition}

\begin{fact}\label{fact:immediately_propagate}
When no failures occur, server actions, such as processing read,
  write, and propagation messages, are executed atomically.
\end{fact}

\begin{fact}\label{fact:fifo}
The network transfer between servers follows a first-in-first-out
  (FIFO) order: if server $S_{i}$ sends messages $m_1$ and then
  $m_2$ to server $S_{j}$, then $m_1$ will be received by $S_{j}$
  before $m_2$.
\end{fact}

\begin{fact}\label{fact:deps_construction}
Based on function ClientWrite in Figure~\ref{fig:alg-server}, when
  processing a write message $w$, the server constructs a version
  $v$ by setting $v.deps = w.deps \cup w.local$, and computing
  $v.vc$ through the following steps: (i) merge vector clocks from
  $w.deps$ and $w.local$ into the server's vector clock $S.vc$,
  then (ii) increment $S.vc$.
Therefore, every $(k, vc)$ pair in $v.deps$ satisfies $vc < v.vc$.
\end{fact}

\heading{Step 1: Propagation safety}
The first group of lemmas establishes that two-round propagation
  ensures all relevant versions reach every server.
The key insight is that traversing a chain twice
  guarantees a complete sub-chain starting from any server
  (Lemma~\ref{lemma:two_rounds_propagate}).
When you combine this insight with the fact that all messages in \sys follow FIFO ordering, this means earlier versions are ``dragged along'' by later ones (Lemma~\ref{lemma:propagates_implies_all_previous_versions_met}).

\begin{lemma}\label{lemma:propagated}
\textit{
When a version $v$ has been propagated successfully, it is
  \globalmet.
}
\end{lemma}

\noindent\textit{Proof.}
A version is considered successfully propagated when it has been
  propagated along a complete chain twice.
When a server receives the propagation message of $v$, $v$ is
  inserted into its \black{}
  (Figure~\ref{fig:alg-server}).
Therefore, for every server $S$, $v$ is present in $S.$\black{}
  and satisfies the condition of \globalmet.
\qed

\begin{lemma}\label{lemma:two_rounds_propagate}
\textit{
When a version $v$ has been propagated successfully, in its
  propagated path, for any cache server $S$, there is a complete
  chain starting from $S$.
}
\end{lemma}
\noindent\textit{Intuition:}
With only one round, a complete sub-chain can only be found
  starting from the head of the chain.
Two rounds ensure that a complete sub-chain exists starting from
  \emph{any} server on the path.
This property is essential for
  Lemma~\ref{lemma:propagates_implies_all_previous_versions_met},
  where an earlier version may originate from a server other than
  the head.

\noindent\textit{Proof.}
By definition, successful propagation means that $v$ has been
  forwarded along a complete chain \emph{twice}, in the forward
  order.
Let a complete chain be denoted as $[S_0, S_1, \dots, S_{n-1}]$,
  where $n$ is the total number of cache servers.
Then, the propagation path includes at least the sequence:
\[
[S_0, S_1, \dots, S_{n-1}, S_0, S_1, \dots, S_{n-1}].
\]
Now, consider any server $S_i$ that appears in this path.
The suffix of the propagation path starting at the first occurrence
  of $S_i$ is:
\[
[S_i, S_{i+1}, \dots, S_{n-1}, S_0, \dots, S_{i-1}, S_i, \dots,
  S_{n-1}],
\]
which forms a complete chain starting from $S_i$ and ending at
  $S_{i-1}$.
Therefore, for any server $S$ on the propagation path, there exists
  a subsequence of the path that forms a complete chain starting
  from $S$.
Hence, the lemma holds.
\qed

\begin{lemma}\label{lemma:propagates_implies_all_previous_versions_met}
\textit{
When a version $v$ has been propagated successfully, then for any
  version $v^{\prime}$ with $v'.vc < v.vc$, $v^{\prime}$ is
  \globalmet.
}
\end{lemma}
\noindent\textit{Intuition:}
FIFO channels ensure that earlier-issued versions arrive before
  later ones.
Combined with Lemma~\ref{lemma:two_rounds_propagate}, a completed
  version ``drags'' all earlier versions to every server.

\noindent\textit{Proof.}
Since version numbers (vector clocks) are assigned when processing
  write messages, and the vector clock increases monotonically, the
  write corresponding to $v^{\prime}$ must have been issued and
  processed by a cache server $S$ before $v$ was issued.
By Fact~\ref{fact:immediately_propagate}, upon receiving the write
  for $v^{\prime}$, server $S$ immediately propagates it to its
  successor.
Thus, $v^{\prime}$ begins its propagation before $v$.

Now, since $v$ has been propagated successfully, it has traversed a
  complete chain twice.
By Fact~\ref{fact:fifo}, messages sent earlier are received earlier.
Therefore, in every server along the propagation path of $v$ that
  starts from $S$, the propagation message for $v^{\prime}$ must
  have arrived before the propagation message for $v$.

Furthermore, by Lemma~\ref{lemma:two_rounds_propagate}, in the
  propagation path of $v$, there exists a complete chain starting
  from $S$.
Hence, $v^{\prime}$ has been propagated to all cache servers.
Therefore, $v^{\prime}$ is \globalmet.
\qed

\heading{Step 2: Integration safety}
The second group of lemmas shows that integration is always safe:
  the dependencies being integrated are \globalmet{} and can be
  found locally, so \white{} remains a causal cut.
The challenge is that the dependencies appearing in integration
  come from two sources, server-side propagation and client-side
  $deps$, and these two sources depend on each other: a client's
  $deps$ are derived from prior reads served from some server's
  \white{}, and integration at a server requires the versions in
  $deps$ to be \globalmet{}.
Establishing integration safety therefore requires a set of
  mutually dependent invariants, which we prove inductively.
  
To reason about which versions are safe to integrate at any given
  point, we introduce a global ghost variable $PVC$ that tracks
  the propagation state of the entire system (used only in the
  proof, not part of the protocol).
Each time a version completes propagation, its vector clock is
  merged into $PVC$.
$PVC$ thus acts as a monotonically growing, system-wide
  ``safety watermark'': any version with a vector clock at or
  below $PVC$ is guaranteed to be \globalmet.

\begin{lemma}\label{lemma:versions_before_pvc_are_met}
\textit{
For any version $v$, if $v.vc \leq PVC$, then $v$ is \globalmet.
}
\end{lemma}
\noindent\textit{Intuition:}
Any version below $PVC$ must have been ``covered'' by some
  completed propagation, and is therefore \globalmet{} by
  Lemma~\ref{lemma:propagates_implies_all_previous_versions_met}.

\noindent\textit{Proof.}
We prove this lemma by induction.

\initial{} $PVC$ is set to the zero vector, and no version has been
  issued.
Thus, the lemma holds.

\inductive{}
Suppose the lemma holds before a server $S$ updates $PVC$ by merging
  the vector clock of a successfully propagated version $v'$.

By Lemma~\ref{lemma:propagated}, $v'$ is \globalmet{}.
Moreover, by
  Lemma~\ref{lemma:propagates_implies_all_previous_versions_met},
  any version $v$ with $v.vc < v'.vc$ is also \globalmet{}.
It follows that all versions $v$ with $v.vc \leq v'.vc$ are
  \globalmet{}.

By the induction hypothesis, all versions $v$ with $v.vc \leq PVC$
  are \globalmet{}.

After the merge, the new $PVC'$ is calculated by taking the
  component-wise maximum of $PVC$ and $v'.vc$.
Therefore, for any version $v$ such that $v.vc \leq PVC'$, it must
  be that either $v.vc \leq PVC$ or $v.vc \leq v'.vc$.
In both cases, $v$ is \globalmet{}.

Thus, the lemma continues to hold after the update on $PVC$.
\qed

\begin{lemma}\label{lemma:deps_are_met}
\textit{
During integration, for any $(k, vc)$ pair in $deps$ to be
  integrated, it holds that $vc \leq PVC$, and the corresponding
  version is \globalmet.
}
\end{lemma}
\noindent\textit{Intuition:}
In both triggering cases (propagation completion and client read),
  the deps are \globalmet{}:
  in the propagation case, they are bounded by $PVC$ since the
  version has completed propagation;
  in the read case, they originate from the client's earlier
  reads, which are \globalmet{} by
  Lemma~\ref{lemma:read_deps_are_met}.

\noindent\textit{Proof.}
Integration is triggered in two cases: \textit{(i)} when a
  propagation message is received, and the version in the message is
  successfully propagated; \textit{(ii)} when a read message is
  received.
In both cases, the message carries a dependency set $deps$.

For case \textit{(i)}, let the propagated version be $v$,
  produced by a corresponding write message $w$.
By Fact~\ref{fact:deps_construction},
  $v.deps = w.deps \cup w.local$.
Since $v$ is successfully propagated, and all versions in $w.local$
  happen before $v$ with a smaller vector clock, by
  Lemma~\ref{lemma:propagates_implies_all_previous_versions_met}
  all versions in $w.local$ are \globalmet.
Then for $w.deps$, by
  Lemma~\ref{lemma:write_deps_are_met} (proved later), each
  $(k, vc)$ pair in $w.deps$ satisfies $vc \leq PVC$ and the
  corresponding version is \globalmet.
Therefore, this lemma holds for the dependency set $v.deps$.

For case \textit{(ii)}, integration is triggered by receiving a
  read message $r$ from the client.
By Lemma~\ref{lemma:read_deps_are_met} (proved later), each
  $(k, vc)$ pair in $r.deps$ satisfies $vc \leq PVC$ and the
  corresponding version is \globalmet.

Overall, once integration happens, all $(k, vc)$ pairs in the
  $deps$ satisfy $vc \leq PVC$, and are therefore \globalmet.
\qed

\begin{lemma}\label{lemma:ccache_is_met}
\textit{
At any time, for each server $S$, every version $v$ in
  $S.$\textit{\white{}} satisfies $v.vc \leq PVC$ and $v$ is
  \globalmet.
}
\end{lemma}
\noindent\textit{Intuition:}
\white{} is only modified by integration, and integration only moves objects with versions that are \globalmet{}.
As a result, everything that is stored in \white{} is safe.

\noindent\textit{Proof.}
We prove this lemma by induction.

\initial{} When all servers have an empty \white{} this lemma trivially holds.

\inductive{}
The \white{} is changed only when integration happens.
Assume the lemma holds before integration.
Let $S$ and $S'$ denote the server's state before and after
  integration.
We need to show that all versions in $S'.$\white{} satisfy the
  lemma.

As described in \S\ref{s:pull}, the integration process begins by
  iterating over the given $deps$ and recursively discovering all
  their transitive dependencies.
According to Lemma~\ref{lemma:deps_are_met}, all versions in the
  initial $deps$ are \globalmet, with each $(k, vc)$ pair satisfying
  $vc \leq PVC$.
Based on Fact~\ref{fact:deps_construction}, every transitive
  dependency of a $(k, vc)$ pair has a vector clock smaller than
  $vc$, and thus is also bounded by $PVC$.
Therefore, all versions to be integrated are \globalmet{} and have
  vector clocks no greater than $PVC$, and the lemma holds after
  integration.
\qed

\begin{lemma}\label{lemma:read_reply_is_met}
\textit{
At any time, for any read reply ($rep$) in the environment,
  $rep.vc \leq PVC$ and the returned version is \globalmet.
}
\end{lemma}
\noindent\textit{Intuition:}
Since reads are served from \white{}, the \globalmet{} property
  established on \white{} by Lemma~\ref{lemma:ccache_is_met}
  carries over to every read reply.

\noindent\textit{Proof.}
We prove this lemma by induction.

\initial{} There are no read replies in the environment, this lemma
  holds.

\inductive{}
Assume the lemma holds before a new read reply is sent.
When a read reply $rep$ is added to the environment, it must have
  been produced by a server $S$ in response to a read request.

The read is served by \white{}, thus the reply is generated from
  $S.$\white{}.
By Lemma~\ref{lemma:ccache_is_met}, every version in $S.$\white{}
  has $vc \leq PVC$ and is \globalmet.
Therefore, the version returned in $rep$ also satisfies
  $vc \leq PVC$ and is \globalmet.
Hence, the lemma continues to hold after the new reply is sent.
\qed

\begin{lemma}\label{lemma:clients_are_met}
\textit{
At any time, for each client $C$, and each $(k, vc)$ pair in
  $C.deps$, $vc \leq PVC$ and the corresponding version is
  \globalmet.
}
\end{lemma}
\noindent\textit{Intuition:}
$C.deps$ only grows by merging in read replies, so the
  \globalmet{} property of replies from
  Lemma~\ref{lemma:read_reply_is_met} propagates to the client's
  dependency map.

\noindent\textit{Proof.}
We prove this lemma by induction.

\initial{}
All clients' $deps$ are empty, this lemma holds.

\inductive{}
A client's $deps$ is updated only upon receiving a read reply.
When client $C$ reads $k$, and receives a reply with $vc$, it
  merges $(k, vc)$ into its $deps$.

By Lemma~\ref{lemma:read_reply_is_met}, the version $vc$ in the
  reply satisfies $vc \leq PVC$ and is \globalmet.
Therefore, the updated $deps$ also satisfy the lemma.
Hence, the lemma holds at all times.
\qed

\begin{lemma}\label{lemma:read_deps_are_met}
\textit{
At any time, for any read request in the environment, every
  $(k, vc)$ pair in its $deps$ satisfies $vc \leq PVC$, and the
  corresponding version is \globalmet.
}
\end{lemma}
\noindent\textit{Intuition:}
Read and write requests copy their $deps$ from the client's
  dependency map.
Note that Lemma~\ref{lemma:deps_are_met} and this lemma are
  mutually dependent; we resolve this by step-wise mutual
  induction.

\noindent\textit{Proof.}
We prove this lemma by induction.

\initial{}
There are no read requests in the environment, this lemma holds.

\inductive{}
Consider a step in which some client $C$ issues a new read
  request into the environment.
The request's $deps$ is copied from $C.deps$
  (Figure~\ref{fig:alg-client}).
By Lemma~\ref{lemma:clients_are_met}, every $(k, vc)$ in $C.deps$
  satisfies $vc \leq PVC$ and is \globalmet.
Therefore, the newly added read request also satisfies the lemma.

Note that this lemma transitively depends on
  Lemma~\ref{lemma:deps_are_met}, which also depends on this lemma.
To handle this mutual induction, we reason step-by-step: assuming
  Lemma~\ref{lemma:read_deps_are_met} holds at step $i$, we can
  prove Lemma~\ref{lemma:deps_are_met} at step $i+1$; conversely,
  assuming Lemma~\ref{lemma:deps_are_met} holds at step $i$, we can
  prove this lemma at step $i+1$.
Therefore, both lemmas hold inductively at every step.
\qed

\begin{lemma}\label{lemma:write_deps_are_met}
\textit{
At any time, for any write request that has been issued, every
  $(k, vc)$ pair in its $deps$ satisfies $vc \leq PVC$, and the
  corresponding version is \globalmet.
}
\end{lemma}
\noindent\textit{Proof.}
This lemma follows by the same argument as
  Lemma~\ref{lemma:read_deps_are_met}.
\qed

\begin{lemma}\label{lemma:ccache_is_causal_cut}
\textit{
At any time, for each server $S$, its \white{} is a causal cut.
}
\end{lemma}
\noindent\textit{Intuition:}
Integration recursively pulls in all transitive dependencies, so
  no ``holes'' are left in \white{}.
Every version's causal predecessors are also present.

\noindent\textit{Proof.}
We prove this lemma by induction.

\initial{}
All servers' \white{}s are empty, this lemma holds.

\inductive{}
Suppose that at some step, a server $S$'s \textit{\white{}}
  changed, which can only occur through integration.
By Lemma~\ref{lemma:deps_are_met}, every version $(k, vc)$ in the
  $deps$ to be integrated is already either in $S.$\white{} or in
  $S.$\black{}.
During integration, all such versions, along with their transitive
  dependencies, are recursively added to $S.$\white{}.
By Fact~\ref{fact:deps_construction}, any transitive dependency of a
  version in $deps$ must precede it.
These dependencies are also either already in $S.$\white{} or will
  be integrated from $S.$\black{}.
Therefore, after integration, for every version in $S.$\white{}, all
  its dependencies are also in $S.$\white{}.
Hence, $S.$\white{} remains a causal cut.
\qed

\heading{Step 3: Client-side guarantees}
All preceding lemmas are machine-checked in
  Dafny~(\S\ref{subsec:dafny_verification}).
We then show how these lemmas, taken together, ensure that clients
  observe causally consistent results.
While the following properties are not themselves machine-checked,
  they follow naturally once all lemmas hold.

From the client's perspective, causal consistency can be modeled by
  the following four properties.

\begin{theorem}[Read Your Writes]
\textit{
If a client performs a write, it can later observe its own writes.
}
\end{theorem}
\noindent\textit{Proof.}
According to Figure~\ref{fig:alg-client}, when a client reads a
  key, if this key exists in its $local$, it will merge the
  returned version with its own writes; therefore, this theorem is
  naturally satisfied.

\begin{theorem}[Monotonic Read]
\textit{
If a client observes a version $(k, vc)$, for its following read on
  $k$, it must observe the same or newer version.
}
\end{theorem}
\noindent\textit{Proof.}
Once the client observes version $(k, vc)$, it merges it into its
  $deps$.
For any subsequent read on key $k$, the client attaches its $deps$
  to the read request.

According to Lemma~\ref{lemma:read_deps_are_met}, all versions in
  $deps$ are \globalmet{}.
Then, after integration, when a server serves this read, all
  versions in $deps$ have been merged into its \white{}.
Therefore, the returned version must have vector clock
  $vc' \geq vc$.

\begin{theorem}[Writes Follow Reads]\label{theorem:write_follow_read}
\textit{
If a client observes $(k_1, vc_1)$ and then performs a write on
  $k_2$, any client that later observes this write on $k_2$ will
  also observe a version of $k_1$ at or newer than $vc_1$.
}
\end{theorem}

\begin{theorem}[Monotonic Writes]\label{theorem:monotonic_writes}
\textit{
If a client performs a write on $k_{1}$ followed by a write on
  $k_{2}$, then any client that observes the write on $k_{2}$ will
  also observe the write on $k_{1}$ or a newer version.
}
\end{theorem}
\noindent\textit{Proof.}
We prove both theorems by following the causal dependency chain.
In both cases, a write on $k_2$ includes $(k_1, vc_1)$ in its
  $deps$: for Theorem~\ref{theorem:write_follow_read}, $(k_1,
  vc_1)$ enters $deps$ as a previously observed version; for
  Theorem~\ref{theorem:monotonic_writes}, it comes from the
  client's own prior write.

Suppose another client $C_2$ later observes this write on $k_2$,
  receiving version $(k_2, vc_2)$.
This version is added to $C_2.deps$.
When $C_2$ reads $k_1$ from some server $S$, the read request
  includes $(k_2, vc_2)$ in its $deps$.
By Lemma~\ref{lemma:read_deps_are_met}, $(k_2, vc_2)$ is available
  at $S$.
After integration, it has been merged into $S.$\white{}.
By Lemma~\ref{lemma:ccache_is_causal_cut}, $S.$\white{} remains a
  causal cut after integration, so all dependencies of
  $(k_2, vc_2)$, including $(k_1, vc_1)$, have also been merged
  into $S.$\white{}.
Therefore, $C_2$ can observe $(k_1, vc')$ with $vc' \geq vc_1$ as
  required in both cases.

\paragraph{Summary.}
Through a series of lemmas, we have shown that server \white{}s are
  always causal cuts, client-tracked dependencies are \globalmet,
  and replies returned to clients respect causal order.
These results together guarantee that any client can always read a
  causally consistent view from any server.

\subsection{Formal Verification in Dafny}\label{subsec:dafny_verification}

We modeled \sys{} and encoded all lemmas from
  \S\ref{subsec:proof} in
  Dafny~\cite{leino2010dafny}.
Dafny provides built-in support for predicates, preconditions,
  postconditions, and loop invariants, and allows developers to
  write lemmas that structure step-by-step reasoning about program
  behavior.
Our encoding focuses on server-side properties that guarantee all
  dependencies in each read request are satisfied on every cache
  server, thereby ensuring that returned values respect causal
  dependencies.
Theorems concerning client-side properties were not encoded, as
  they follow directly once all lemmas hold.
Altogether, the protocol specification and lemma proofs comprise
  approximately 7,500 lines of Dafny code.

\heading{Verification challenges}
Two aspects of the verification were particularly challenging.

\weakheading{Finding inductive invariants}
The central difficulty in deductive verification is identifying
  invariants that are simultaneously strong enough to imply the
  desired correctness properties and inductive so that every system
  action preserves them.
For \sys{}, the key design decision was introducing the ghost
  variable $PVC$ (\S\ref{subsec:proof}), which serves as the
  bridge between propagation completion and version-level safety.
Arriving at this formulation was non-trivial: without $PVC$,
  relating propagation completion to safety would require
  case-splitting over all possible propagation orderings, which is an approach that does not scale.

\weakheading{Proving action preservation}
Once we identify the invariants, we must show that every action of the protocol (e.g., client reads, writes,
  propagation, integration) preserves them.
This produces a large number of proof obligations that require manual guidance.
For example, the integration-related lemmas involve recursive
  traversal of transitive dependencies, where each recursive step
  must be justified against the invariants.
A further complication is that our invariants quantify over all
  servers, versions, and dependencies at once, and Dafny's
  underlying solver slows down dramatically when many such
  quantifiers interact.
We therefore had to carefully organize the proof so that the
  solver only reasons about the relevant portion of the state at a time, rather than unfolding every invariant everywhere.

\heading{Lessons learned}
First, model checking and deductive verification are
  complementary rather than interchangeable.
Model checking (e.g., TLA+/TLC) is effective for rapidly
  debugging a protocol during design iterations, but as
  \S\ref{subsec:violation} illustrates, it alone cannot guarantee
  correctness at arbitrary scales.
When building foundational system infrastructure, deductive
  verification is worth the additional effort to obtain the
  strongest correctness guarantees.
Second, when a proof attempt persistently fails, the stuck proof
  obligation often points to a flaw in the protocol itself rather
  than a gap in proof technique, and is worth investigating as a
  diagnostic signal.
Finally, in deductive verification, the design of auxiliary
  ghost state (such as $PVC$ in our proof) is where the most
  creative leverage lies: the right abstraction can turn an
  otherwise intractable proof into a tractable one.

\section{Related Work}\label{s:relwork}
\topheading{Causal consistency}
There is a large body of work on causally consistent systems
  \cite{petersen1997bayou,yu2000design,belaramani2006practi,lloyd2011cops,du2013orbe,bailis2013bolt,almeida2013chainreaction,lloyd2013Eiger,du2014gentlerain,zawirski2015writefast,akkoorath2016cure,mehdi2017occult,tyulenev2019mongodb}.
Several systems (such as Bolt-on~\cite{bailis2013bolt},
  SwiftCloud~\cite{zawirski2015writefast}, and Occult~\cite{mehdi2017occult}) 
  have explored the idea of reading safe but stale data to achieve causally 
  consistent plus (CC+) guarantees.
However, these systems are not designed to guarantee CC+ across multiple caches. 
They either do not support client roaming or if one deploys them
  in a setting with client roaming they must block when dependencies are not 
  satisfied. 
As a result, they are not suitable for serverless computing where mobility is a 
  common case.
\sys, on the other hand, is specifically designed to support client migration and guarantee CC+ across multiple caches.
Moreover, \sys{}'s correctness has been machine-checked via Dafny
  for arbitrary system sizes, which to our knowledge is the first such
  verification for a causally consistent cache system.

Chain Reaction~\cite{almeida2013chainreaction} introduces the idea of using 
  chain replication for causal consistency, but we have applied this concept 
  in a different manner. 
In their single\hyp{}datacenter setting, keys are sharded to different chains 
  using consistent hashing, and reads are forwarded to the server that holds 
  the key. 
In their multi-datacenter setting, the heads of the chains in different 
  datacenters are connected. 
If the required version is missing, the request will be forwarded to other 
  datacenters or blocked until the version becomes available. 
In contrast, \sys{} allows a server to serve requests without coordinating 
  with other servers, and the chains are interleaved (one's head to another's tail) 
  rather than parallel (one's head to another's head), which enables 
  coordination-free reads.

\heading{Serverless caching}
Other systems have also noted the high cost of remote data access in serverless computing.
HydroCache~\cite{wu2020hydrocache} and FaaSTCC~\cite{lykhenko2021faastcc} are the
  most closely related works.
One difference between \sys{} and these works is that \sys{} is non-blocking and 
  has advantages such as supporting fully dynamic workflows~(\S\ref{s:discuss}),
  while these works must block when a specific version is missing, which
  nullifies many of the benefits of caching.
In terms of requirements, \sys{} and HydroCache can run on top of existing databases, whereas
FaaSTCC needs the underlying storage to assign a causal 
timestamp for each write, which is not supported by most production databases.

Also related is Faa\$T~\cite{romero2021faat}, which offers strong consistency but requires validating the version at the remote storage to ensure that the cached data is up to date, introducing high latency.
Cloudburst~\cite{sreekanti2020cloudburst} is a serverless platform that supports 
  a causal cache by using lattice data types provided by 
  Anna KVS~\cite{wu2019anna,wu2019anna2} with its custom API.\@
Other ephemeral storage and caches, like Pocket~\cite{klimovic2018pocket}, 
InfiniCache~\cite{wang2020infinicache}, and Locus~\cite{pu2019shuffling} are 
  designed for data-intensive serverless apps like data analytics 
  and lack consistency guarantees. 

\heading{Serverless scheduling and orchestration}
Finally, we note that \sys{} is complementary to recent work on more 
  efficient scheduling for serverless computing (e.g., Kaffes~\cite{kaffes2019centralized}, 
  Fifer~\cite{gunasekaran2020fifer}, Pheromone~\cite{yu2021pheromone},
  Unum~\cite{liu2023unum}, Cypress~\cite{bhasi2022cypress},
  Hermod~\cite{kaffes2022hermod}, Palette~\cite{abdi2023palette},
  and Caerus~\cite{zhang2021caerus}).
  In particular, \sys{} adds an extra layer to the efficiency of function 
    placement that scheduling algorithms can leverage to improve the performance 
    of workflow execution while ensuring causal consistency.
Integration may differ slightly for different schedulers, but we leave a 
  full exploration of the optimal co-design of the end-to-end caching, 
  instance provisioning, and request scheduling infrastructure to future work.

\section{Discussion}\label{s:discuss}

\heading{Dynamic workflows}
Recent analyses~\cite{luo2021alibaba,huye2023lifting} show that workflows and operation sets tend to be highly dynamic.
For example, consider a function that reads the value of key $k_1$ and then uses the value of $k_1$ as the key for a subsequent read operation.
HydroCache cannot support this type of function.
HydroCache-Con only supports static workflows because it requires knowledge of the read set before execution. HydroCache-Opt can support partially dynamic workflows, but it still requires knowledge of the read set of each function before execution and then performs a validation phase to check for causal violations. In contrast, both \sys{} and \sys{}-TCC support arbitrary dynamic workflows. 

\heading{Metadata and garbage collection}
In dependency-tracking systems, the accumulation of dependency metadata can cause system slowdown over time.
To mitigate this issue, HydroCache implements periodic garbage collection (GC) using a background consensus protocol to clear the dependency metadata. In contrast,
\sys{} and \sys-TCC clear unnecessary metadata seamlessly while processing requests, without the need for dedicated GC processes. Specifically, in \sys{}, dependencies are discarded during dependency integration. In \sys{}-TCC, \white{} is a ring buffer that automatically removes both old values along with their associated dependencies when it is full. 

\heading{Scale to more cache nodes}\label{s:discuss:scale}
The usage of vector clocks can potentially lead to performance issues if the vector clock becomes huge, e.g., supporting over 1000 cache nodes. However, it is generally not encountered in serverless setups where multiple workers can be routed to the same cache server. 
For example, DynamoDB DAX is designed to allow up to 11 cache nodes; CosmosDB's Integrated Cache allows a maximum of 5 cache nodes.
Should the system go to an extreme scale in the future and the vector clock becomes a bottleneck, we expect to utilize a garbage collection scheme to practically trim the vector clock, such as Dynamo did~\cite{decandia2007dynamo}. 

\heading{Failure recovery}
If a server fails and propagation cannot complete, the
  fault-tolerant coordinator reconfigures the chain by removing
  the failed server and restarting the remaining servers, after
  which normal propagation resumes.
During reconfiguration, serverless functions read directly from
  the persistent database. As \sys{} persists all writes before returning to the clients, upon failure, the cache can be safely dropped without losing writes.

\heading{Cache eviction strategies} 
Cache eviction is an orthogonal problem and thus is not the focus of our work. The design of dual cache allows it to benefit from any eviction policies. The only additional requirement is upon the eviction of a key, all keys that depend on it are also evicted so that \white{} remains a cut.

\section{Conclusion}
This paper presents \sys{}, the first cache system to support
  coordination-free and abort-free causal read/write operations
  when clients (workflows) move from server to server, and
  \sys{}-TCC, a variant that supports transactional causal
  consistency.
Both protocols enable developers to build applications that
  combine the scalability of serverless computing with the low
  latency of a local cache.
Our evaluation shows that \sys{}(-TCC) significantly outperforms the state-of-the-art consistent caches, and 
\sys{} is the only one to be formally verified.

\section*{Source code}

Our code is open source and available at:
\begin{center}
    \url{https://github.com/eniac/causalmesh}
\end{center}

\section*{Technical perspective}
Phil Bernstein wrote a nice technical perspective on CausalMesh~\cite{bernstein25causal}.

\bibliographystyle{spmpsci}      %
\bibliography{conferences,paper}   %

\end{document}